\documentclass{aastex62}

\received{\today}
\revised{\today}
\accepted{\today}

\submitjournal{ApJS}

\begin{document}

\title{Distribution Properties of the 6.7 GHz Methanol Masers and Their Surrounding Gases in the Milky Way}

\correspondingauthor{Xi Chen}
\email{chenxi@gzhu.edu.cn}

\author{Tian Yang}
\affiliation{Center for Astrophysics, Guangzhou University, Guangzhou 510006, People’s Republic of China}

\author{Xi Chen}
\affiliation{Center for Astrophysics, Guangzhou University, Guangzhou 510006, People’s Republic of China}
\affil{Shanghai Astronomical Observatory, Chinese Academy of Sciences, 80 Nandan Road, Shanghai 200030, People’s Republic of China}

\author{Yan-Kun Zhang}
\affiliation{Center for Astrophysics, Guangzhou University, Guangzhou 510006, People’s Republic of China}

\author{Xu-Jia Ouyang}
\affiliation{School of Physics and Astronomy, Sun Yat-sen University, 2 Daxue Road, Tangjia, Zhuhai, Guangdong Province, People’s Republic of China}

\author{Shi-Min Song}
\affiliation{Center for Astrophysics, Guangzhou University, Guangzhou 510006, People’s Republic of China}

\author{Jia-Liang Chen}
\affiliation{Center for Astrophysics, Guangzhou University, Guangzhou 510006, People’s Republic of China}

\author{Ying Lu}
\affiliation{Center for Astrophysics, Guangzhou University, Guangzhou 510006, People’s Republic of China}

\begin{abstract}
An updated catalog consisting of 1092 6.7-GHz methanol maser sources was reported in this work. Additionally, the NH$_3$ (1, 1), NH$_3$ (2, 2), and NH$_3$ (3, 3) transitions were observed towards 214 star forming regions using the Shanghai Tianma radio telescope (TMRT) in order to examine the differences in physical environments, such as excitation temperature and column density of molecular clouds associated with methanol masers on the Galactic scale. Statistical results reveal that the number of 6.7 GHz methanol masers in the Perseus arm is significantly lower than that in the other three main spiral arms. In addition, the Perseus arm also has the lowest gas column density among the main spiral arms traced by the NH$_3$ observations. Both of these findings suggest that the Perseus arm has the lowest rate of high-mass star formation compared to the other three main spiral arms. We also observed a trend in which both the luminosity of 6.7 GHz methanol masers and the ammonia gas column density decreased as the galactocentric distances. This finding indicates that the density of material in the inner Milky Way is generally higher than that in the outer Milky Way. It further suggests that high-mass stars are more easily formed at the head of spiral arms. Furthermore, we found that the column density of ammonia gas is higher in the regions on the arms than that in the inter-arm regions, supporting that the former is more likely to be the birthplace of high-mass stars.
\end{abstract}

\section{Introduction} \label{sec:1}
Star formation is a fundamental and important issue in modern astronomy, closely linked to the origin and evolution of cosmic objects at all scales. In particular, the formation of high-mass stars (M $>$ 8 M$_\odot$) plays a crucial role in the structure and evolution of galaxies. However, research on high-mass star formation is limited by the scarcity of observable samples of high-mass young stellar objects (HMYSOs) and their short evolutionary timescale. Furthermore, the far distance of these objects and the optical obscuration caused by molecular clouds severely hinder observations \citep{2007ARA&A..45..481Z,2018ARA&A..56...41M}. Despite these challenges, high-mass star forming regions (HMSFRs) can still be investigated through various observational tracers, such as interstellar molecules and masers. In particular, interstellar masers, with their extremely high spectral intensity, serve as important probes in investigating HMSFRs \citep{1981ARA&A..19..231R}.

Methanol masers are frequently observed toward HMSFRs within the Milky Way in radio band. They are classified into two types based on their associations with various star formation phenomena. Class I methanol masers, such as the 36.2 and 95.2 GHz methanol masers, are pumped by collisions in shock or outflow regions \citep{2011ApJS..196....9C,2013ApJS..206....9C,2016A&A...592A..31L}. On the contrary, Class II methanol masers which include the 6.7 and 12.2 GHz ones, are typically observed in close proximity to the HMSFRs and are pumped by infrared radiation from the HMSFRs \citep{2005MNRAS.360..533C}.

The 5$_1$ -- 6$_0$ A$^+$ maser emission at 6.668 GHz is the most widely distributed and is detected only in HMSFRs, making it a unique indicator of HMSFRs \citep{2003A&A...403.1095M,2003ChJAA...3...49X}. It was first discovered by \citet{1991ApJ...380L..75M}, and since then, numerous targeted and unbiased surveys have been conducted over the past three decades. Targeted surveys aim to discover 6.7 GHz methanol masers from existing HMSFR candidates identified by other tracers such as infrared sources, H$_2$O masers, and OH masers \citep{1991ApJ...380L..75M,1992MNRAS.254P...1M,1992MNRAS.256..519M,1993MNRAS.261..783S,1993MNRAS.262...43G,1995A&AS..110...81V,1995MNRAS.272...96C,1996MNRAS.282.1085V,1997MNRAS.291..261W,1998AJ....116.2936M,1999A&AS..134..115S,2000A&AS..143..269S}. In contrast, unbiased surveys detect all sources in a specific region of the sky, limited only by the sensitivity of observation \citep{1996MNRAS.279...79C,1996MNRAS.280..378E,2000A&A...362.1093M,2002A&A...384L..15P,2002A&A...392..277S}. \citet{2005A&A...432..737P} compiled a catalog of 6.7 GHz methanol masers, totaling 519 sources. While the detection of 6.7 GHz methanol masers in the Milky Way is still incomplete, some further surveys have been carried out to enrich this 6.7 GHz methanol maser catalog \citep{2007ApJ...656..255P,2007MNRAS.377..571E,2008A&A...485..729X,2009ApJ...702.1615C,2014A&A...566A..18O,2017ApJ...846..160Y,2019ApJS..241...18Y,2019ApJS..245...12O}, especially the Methanol Multibeam (MMB) survey. The MMB survey utilized the Parkes telescope to conduct a blind survey across a relatively broad area covering the Galactic plane (186$^\circ$ $\leq$ l $\leq$ 60$^\circ$, $|b|$ $\leq$ 2$^\circ$) and identified a total of 972 sources, with some overlaps with sources discovered in previous surveys \citep{2009MNRAS.392..783G,2010MNRAS.409..913G,2012MNRAS.420.3108G,2017MNRAS.469.1383G,2010MNRAS.404.1029C,2011MNRAS.417.1964C,2015MNRAS.450.4109B,2018MNRAS.474.3898B}. To date, over a thousand 6.7 GHz methanol masers have been detected in the Galaxy \citep{2019ApJS..241...18Y}, covering the entire galactic plane.

Since the 6.7 GHz methanol maser is a good target for the parallax measurement and a tracer for properties of HMSFRs, it becomes a valuable tool for measuring the structure of the Milky Way. By accurately locating the HMSFRs wherein the 6.7 GHz methanol masers are produced, we can delineate the spiral arm structure of the Milky Way on a large scale. The Bar and Spiral Structure Legacy (BeSSeL) Survey utilized the Very Long Baseline Array (VLBA) to measure the trigonometric parallax towards a large sample of methanol masers and water masers across the Milky Way \citep{2016ApJ...823...77R}. They have achieved the parallax of approximately 200 maser sources, and used these measurement results to depict the spiral arm structures of the Milky Way and calculate the fundamental galactic kinematic parameters. Based on these measurements, the BeSSeL team suggested that the Milky Way consists of four main spiral arms: the Scutum-Centaurus-OSC arm, the Norma-Outer arm, the Perseus arm, and the Sagittarius-Carina arm. Additionally, the team updated the fundamental constants of the Milky Way, such as the distance between the Sun and the Galactic center, which was revised to 8.15 kpc \citep{2019ApJ...885..131R}. The BeSSeL team also developed a program for estimating kinematic distances, which enables us to obtain updated kinematic distances to nearly all of the 6.7 GHz methanol maser sources.

Ammonia molecules are excellent tracers of dense gases ($\sim$ 10$^4$ cm$^{-3}$), which are commonly associated with the hot core phase of massive star formation \citep{1983ARA&A..21..239H}. In addition, ammonia is excited in gases at kinematic temperatures ($>$ 5 K) and is found in cold (T $<$ 10 K) dense molecular clouds \citep{2012MNRAS.426.1972P}. Maintaining a gas phase in extremely cold regions is a challenging task for common molecular tracers like CO, which tend to freeze to the surface of dust grains. Ammonia molecules, however, can maintain a gas phase at low temperatures, being more resistant to depletion \citep{2011MNRAS.418.1689U}. The ammonia J,K = (1,1) inversion transition displays a significant hyperfine structure that can be utilised to deduce the optical depth of the molecular cloud. By combining the (1,1) transition with other higher order inversion transitions, it is possible to estimate the rotational temperature of the gas. In addition, the column density of the molecular gas can be determined using the spectral line information of ammonia under local thermodynamic equilibrium. Thus, ammonia molecules can play an important role in the determination of the physical parameters of the gases surrounding the HMSFRs.

There are still questions about the differences between different arms within the Milky Way. For example, how different are the HMSFRs on different arms, and between on-arms and inter-arms? We hope to obtain a more comprehensive understanding to these questions in this work. The primary objective of this paper is to perform a statistical analysis for the HMSFR tracer 6.7 GHz methanol maser in order to examine the distribution characteristics of their locations and luminosities on a large scale within the Milky Way. Additionally, by combining our observations of ammonia gas, we attempt to identify differences of molecular clouds in various environments. This will contribute to a more comprehensive understanding of the overall structure associated with HMSFRs in the Mikly Way.

The rest of the paper is developed as follows. Section \ref{sec:2} describes the compilation of a catalog of 6.7 GHz methanol maser sources and the sample and observation of ammonia transitions using Tianma radio telescope (TMRT) at K-band. The updated distances to the 6.7 GHz methanol masers with their distributions on the Galactic plane, as well as the detection of ammonia transitions are described in Section \ref{sec:3}. In Section \ref{sec:4}, some statistical investigations for the large-scale distributions of methanol masers (including luminosity and distance), and the gas temperature and column density of the molecular clouds traced by ammonia transitions are given. Finally, the conclusion is summarized in Section \ref{sec:5}.

\section{Samples and Observations} \label{sec:2}
\subsection{An updated catalog of the 6.7 GHz methanol masers} \label{subsec:2.1}
A relatively comprehensive catalog of the 6.7 GHz methanol masers with high observation sensitivity has been compiled from the MMB survey on the sky of the southern hemisphere
\citep{2010MNRAS.404.1029C,2011MNRAS.417.1964C,2010MNRAS.409..913G,2012MNRAS.420.3108G,2015MNRAS.450.4109B}. However, the northern sky still lack a comprehensive survey of 6.7 GHz methanol masers. Recently, we used the TMRT to search 3348 infrared sources selected from the Wide-Field Infrared Survey Explorer (WISE), based on specific criteria for infrared luminosity, and a total of 224 6.7-GHz methanol masers were detected, including 32 newly discovered ones (see \citealt{2017ApJ...846..160Y,2019ApJS..241...18Y}). Combining previous survey, a catalog of 1,085 sources was compiled in \citet{2019ApJS..241...18Y}. However, the catalog missed three new 6.7 GHz methanol maser sources discovered by \citet{2017ApJ...846..160Y} at high galactic latitude regions. In addition, \citet{2019ApJS..245...12O} utilized the TMRT to identify four previously unknown 6.7 GHz methanol maser sources. The information from these seven sources is presented in Table \ref{table1}. In conjunction with Table \ref{table1} and the catalog compiled by \citet{2019ApJS..241...18Y}, we have created an updated catalog of 6.7 GHz methanol masers, which includes 1092 sources. Appendix Table \ref{appendix/methanol} summarizes the related maser characteristics of all 1092 maser sources.

\begin{deluxetable*}{crrrrrccl}[h]
\tablecaption{Seven 6.7 GHz methanol maser sources in addition to \citet{2019ApJS..241...18Y} catalog \label{table1}}
\tablewidth{0pt}
\tablehead{
\colhead{Name}              & \colhead{R.A.(J2000)}   & \colhead{Dec.(J2000)}                       & \colhead{V$_l$}         & \colhead{V$_u$}         & \colhead{V$_p$}             & \colhead{S$_p$}         & \colhead{Parallax}                          & \colhead{Ref} \\
\colhead{($^{\circ}$ $^{\circ}$)} & \colhead{(h m s)}       & \colhead{($^{\circ}$ $^{\prime}$ $^{\prime\prime}$)} & \colhead{(km s$^{-1}$)} & \colhead{(km s$^{-1}$)} & \colhead{(km s$^{-1}$)}     & \colhead{(Jy)}          & \colhead{(mas)}                             & \colhead{}
}
\decimalcolnumbers
\startdata
    G017.964$+$00.080& 18 23 20.83& $-$13 15 05.40&  23.5&  28.2&  23.8&   0.80&               &O19    \\
    G019.074$-$00.286& 18 26 48.55& $-$12 26 27.60&  58.0&  72.0&  58.8&   0.90&               &O19    \\
    G026.109$-$00.094& 18 39 17.14& $-$06 06 43.90&  25.4&  27.6&  26.3&   0.40&               &O19    \\
    G097.527$+$03.184& 21 32 11.30& $+$55 53 39.60& -78.4& -67.6& -70.9&   1.00&0.133$\pm$0.017&Y17;R19\\
    G110.196$+$02.476& 22 57 29.80& $+$62 29 45.10& -57.2& -56.4& -56.9&   0.50&0.314$\pm$0.070&Y17;R19\\
    G137.068$+$03.002& 02 58 13.30& $+$62 20 31.90& -59.8& -58.1& -59.0&   1.60&               &Y17    \\
    G236.816$+$01.982& 07 44 27.94& $-$20 08 33.00&  42.0&  49.0&  43.2&   0.80&               &O19    \\
\enddata
\tablecomments{Columns (1) -- (3): source name, and source positions in equatorial coordinates. Columns (4) -- (6): the lower and upper velocities and the peak velocity of the 6.7 GHz methanol maser spectra. Column (7): the maser peak flux density. Column (8): parallax measured by BeSSeL team. Column (9): references -- Y17 \citep{2017ApJ...846..160Y}, O19 \citep{2019ApJS..245...12O}, R19 \citep{2019ApJ...885..131R}.}
\end{deluxetable*}

\subsection{Sample and observation of ammonia}\label{subsec:2.2}
In this work, we selected a total of 214 sources as our sample for the observations of the ammonia transitions. The sample was selected from either previously known ammonia detection or 6.7 GHz methanol maser sources, as shown in Table \ref{table2}. The sample traces different stages of star formation, including infrared dark clouds (IRDCs), HMYSOs, and H{\sc ii} regions \citep{1996A&A...314..265W,2007MNRAS.379..535L,2013ApJ...764...61C,2014ApJ...783..130R}. By cross-matching the observed ammonia sources with the 6.7 GHz methanol masers, out of 214 sources, 129 are identified to have 6.7 GHz methanol masers.

\begin{deluxetable*}{ccrccccrrrrr}
\tablecaption{Sample of ammonia observations}
\tablehead{
\colhead{Name} & \colhead{R.A.(J2000)} & \colhead{Dec.(J2000)} & \colhead{NH$_3$ (1, 1)} & \colhead{NH$_3$ (2, 2)} & \colhead{NH$_3$ (3, 3)} & \colhead{CH$_3$OH} & \colhead{Dist} & \colhead{+/-} & \colhead{Prob} & \colhead{Arm} & \colhead{D$_{GC}$}\\
\colhead{($^{\circ}$ $^{\circ}$)} & \colhead{(h m s)} & \colhead{($^{\circ}$ $^{\prime}$ $^{\prime\prime}$)} & \colhead{} & \colhead{} & \colhead{} & \colhead{} & \colhead{(kpc)} & \colhead{(kpc)} & \colhead{} & \colhead{} & \colhead{(kpc)}
}
\decimalcolnumbers
\startdata
G000.376$+$00.040 & 17 46 21.41 & $-$28 35 40.0 & Y & Y & Y & Y & 8.00  & 3.01 & 1.00 & 3kF & 0.16 \\
G000.409$-$00.504 & 17 48 33.48 & $-$28 50 52.5 & Y & Y & Y & Y & 2.82  & 0.22 & 0.96 & ScN & 5.33 \\
G000.527$+$00.182 & 17 46 09.80 & $-$28 23 29.0 & Y & Y & Y & Y & 2.80  & 0.23 & 0.37 & ScN & 5.35 \\
G000.677$-$00.025 & 17 47 19.28 & $-$28 22 14.8 & N & N & N & Y & 7.87  & 0.90 & 0.11 & ... & 0.30 \\
G000.860$-$00.067 & 17 47 54.71 & $-$28 14 09.9 & N & N & N & N &       &      &      &     &      \\
G002.615$+$00.134 & 17 51 12.30 & $-$26 37 37.2 & Y & Y & Y & Y & 7.92  & 2.13 & 0.86 & ... & 0.43 \\
G002.703$+$00.040 & 17 51 45.98 & $-$26 35 56.7 & Y & Y & Y & Y & 9.90  & 10.29& 1.00 & ... & 1.80 \\
G005.618$-$00.082 & 17 58 44.78 & $-$24 08 40.1 & Y & Y & Y & Y & 4.67  & 0.25 & 0.45 & 3kN & 3.53 \\
G006.368$-$00.052 & 18 00 15.82 & $-$23 28 43.8 & Y & Y & Y & Y & 8.10  & 2.02 & 1.00 & ... & 0.90 \\
G007.601$-$00.139 & 18 03 14.43 & $-$22 27 00.9 & Y & Y & Y & Y & 8.09  & 2.02 & 1.00 & ... & 1.08
\enddata
\tablecomments{Column (1): source name. Columns (2) -- (3): source positions in equatorial coordinates. Columns (4) -- (7): information on whether the source was detected by other tracers: Y -- detected; N -- undetected. Columns (8) and (9): distance and its error. Column (10): probability of being in the spiral arm. Column (11): the spiral arm name of the source location. Column (12): Galactocentric distance. 
(This table is available in its entirety in machine-readable form.)}
\label{table2}
\end{deluxetable*}

Among the 214 sources in our sample, we conducted observations of the (J, K) = (1, 1), (2, 2), and (3, 3) lines of ammonia in April 2019 for 175 sources that had been previously surveyed for ammonia \citep{1996A&A...314..265W,2007MNRAS.379..535L,2013ApJ...764...61C}. Then, we conducted the same ammonia observations in January and April 2020 for the remaining 39 sources associated with 6.7 GHz methanol masers \citep{2014ApJ...783..130R}. These observations were carried out using the TMRT with a beam size of 50$^{\prime\prime}$ at the observed ammonia frequencies. A cryogenically cooled K-band (17.9 -- 26.2 GHz) receiver and the digital backend system (DIBAS) were used to receive and record the spectral line data. Three 23.4 MHz spectral windows with 16,384 channels each were used in the observations to cover the three ammonia transitions, yielding a channel spacing of 1.43 kHz, corresponding to a velocity spacing of $\sim$ 0.02 km s$^{-1}$. The system temperature range is 100--200 K. The main beam efficiency of the telescope in the K-band is about 60\%, and the aperture efficiency is about 50\% (corresponding to the sensitivity of $\sim$1.66 Jy K$^{-1}$). 

Each source observation consists of $\sim6$ repetitions with a 3 minute ON/OFF cycle in position-switching mode, resulting in the total on-source time of $\sim9$ minutes for each. A noise diode was utilized for the prior flux calibration, achieving the flux density uncertainty of less than 10\%.

The ammonia data is processed using the Continuum and Line Analysis Single-dish Software (CLASS) of the Grenoble Image and Line Data Analysis Software packages (GILDAS; \citealt{2005sf2a.conf..721P}). To obtain the final spectra, the data from all scans were first averaged, and a fitted baseline was subsequently subtracted from all channels. In order to enhance the detection of weaker signals, the original spectrum with a lower signal-to-noise ratio (SNR) was smoothed to a velocity space of $\sim0.3$ km s$^{-1}$ based on the line intensity. The finally produced spectra have a typical root mean square (rms) noise of $\sim0.1$ K.

\section{Results} \label{sec:3}
\subsection{Distances and Galactic locations of the 6.7 GHz methanol masers} \label{subsec:3.1}
Ideally, the distance to the maser source can be determined directly measurement of its parallax with Very Long Baseline Interferometry (VLBI). The BeSSeL project has measured the parallax and proper motion toward $\sim$200 maser sources, including 6.7 GHz methanol masers associated with HMSFRs \citep{2019ApJ...885..131R}. By comparing the coordinates of the $\sim$200 maser sources, we identified 143 sources out of the catalog containing 1092 6.7-GHz methanol maser sources to be measured accurate parallax distances. For these sources, their parallax distances are adopted in this work.

For the remaining 949 sources for which no astrometric parallax measurements are available, kinematic distances are used. The main method for estimating the kinematic distances is through the assumption of the Galactic flat rotation curve \citep{2019ApJ...885..131R}. However, sources within the solar circle pose the challenge of the ambiguity of the kinematic distance. This implies that for a given line of sight, the velocity can correspond to two plausible distances -- one situated on the near-side of the tangent distance, and the other one on the far-side. Currently, there are two techniques available for resolving kinematic distance ambiguities: the H{\sc i}SA technique, by examining the spectral profile of the H{\sc i}, and the Bayesian technique, by correlating the maser source with the spiral arm obtained from astrometric parallax measurements. The H{\sc i}SA technique assesses whether the cloud is in the foreground by linking the cold neutral medium (detected by the cold H{\sc i}) with the molecular cloud compared to the emission from the background \citep{2011MNRAS.417.2500G}. If there is absorption in the H{\sc i} spectrum near the systemic velocity of the source, then the cold neutral medium associated with the source is likely to be nearby, as it absorbs into the background H{\sc i} emission in the foreground. In contrast, if it is not detected, it is probably located at a further kinematic distance. The Bayesian technique consists of assigning maser sources to the spiral arm and combining multiple probabilities associated with the arm, kinematic distances and parallax measurements to produce a probability density function for distance \citep{2019ApJ...885..131R}. 


We adopt the latest parallax-based distance calculator\footnote{\url{http://bessel.vlbi-astrometry.org/node/378}} to determine their distances and locations in the Milky Way. This is the most accurate kinematic distance calculator available, where the distance to the galactic center, R$_0$ = 8.15 $\pm$ 0.15 kpc, and the speed of the Sun's circular rotation is $\Theta$ = 236 $\pm$ 7 km s$^{-1}$. In the calculations, the V$_{lsr}$ value was adopted from the peak velocity V$_{p}$ value of the source and the error in V$_{lsr}$ was set to 8 km s$^{-1}$ considering the potential streaming motions. However, the method of \citet{2019ApJ...885..131R} relies on maser measurements taken from the northern hemisphere, and lacks sufficient measurements from the southern hemisphere. Furthermore, the majority of methanol maser sources in the MMB catalog are situated in the galactic plane of the southern hemisphere. Therefore, directly using Bayesian distances may not provide a highly reliable result. Considering that the H{\sc i}SA method has southern hemisphere measurements and that the H{\sc i}SA distances can help us determine whether a solar circle source is located near or far, we refer to the H{\sc i}SA method to determine whether we should choose the near or far Bayesian distance. A total of 778 MMB maser sources were resolved the kinematic distance ambiguities using the H{\sc i}SA method by \citet{2011MNRAS.417.2500G} and \citet{2017MNRAS.469.1383G}. 
Cross-matching the 949 sources without parallax measurements with the H{\sc i}SA distance sources (\citealt{2011MNRAS.417.2500G,2017MNRAS.469.1383G}), yielded 572 sources with H{\sc i}SA distances and 377 sources without H{\sc i}SA distances. For the former 572 sources, we served the H{\sc i}SA distance as a reference, and selected the updated-kinematic distance which is closest to the H{\sc i}SA distance as the source distance from several sets of Bayesian distances. If all sets of Bayesian distances differ significantly from the H{\sc i}SA distance ($>$ 3 kpc), we consider the H{\sc i}SA distance to be not credible, and we directly use the set of Bayesian distances with the highest probability. For the latter 377 sources which have neither parallax nor H{\sc i}SA distances, we use the set of Bayesian distances with the highest probability. The above method will allow us to obtain more reliable kinematic distances.

\begin{figure}[htbp!]
    \plotone{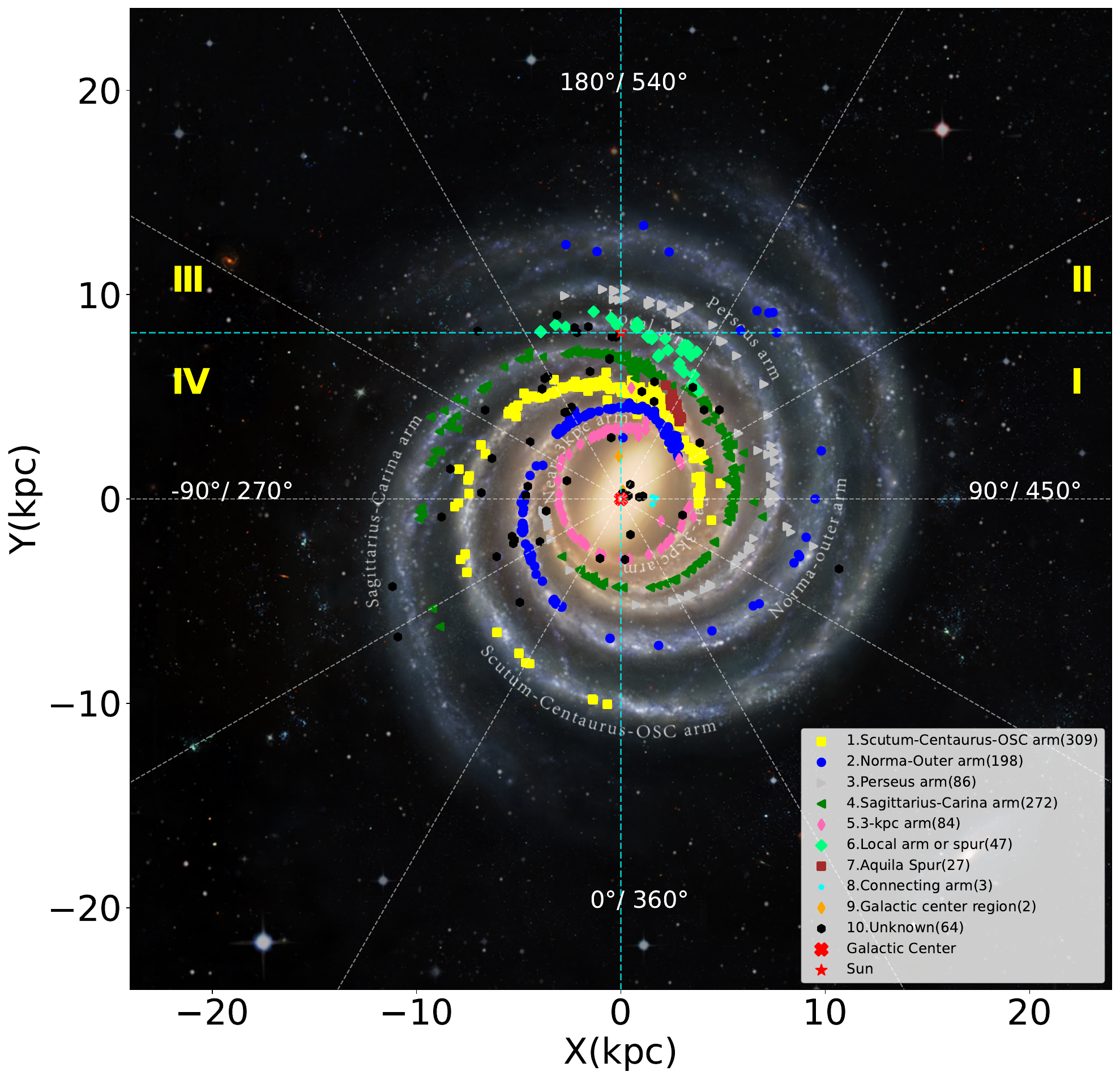}
    \caption{Distribution map of the 6.7 GHz methanol masers in the Galactic plane. The background image is an artistic picture of the Milky Way. The Galactic center is located at (0, 0) kpc, while the Sun is situated at (0, 8.15) kpc. The Milky Way is divided into four quadrants by two mutually orthogonal cyan dashed lines. Each quadrants is represented by yellow Roman numerals. The direction of the Sun towards the Galactic center is set at a spiral angle (i.e., SA) of 0$^{\circ}$, which increases counterclockwise and is indicated by the white dashed lines in our analysis. In the lower right corner, the names of the Galactic structure features and the number of sources located within them are given.}
    \label{fig Overlaying}
\end{figure}

Figure \ref{fig Overlaying} illustrates the distribution of 6.7 GHz methanol masers on the Galactic plane. We categorized all sources into the four main spiral arms, five secondary structures, and an "Unknown" category for sources whose allocations can not be effectively defined at any known structure features of the Milky Way. We then overlapped their positions onto an artistic image of the Milky Way\footnote{\url{https://astronomy.nju.edu.cn/xtzl/EN/index.html}}. As depicted in Figure \ref{fig Overlaying}, the distribution of 6.7 GHz methanol maser sources covers the entire galactic plane, spanning all four quadrants. The continuity of these sources along the same spiral arm extends outward, signifying that they do indeed trace the spiral arms of the Milky Way. Notably, majority of the ``Unknown''  sources on the distribution map are located in the space between the spiral arms. Therefore, we treat these Unknown sources as the inter-arm sources in our analysis. While there are still a few Unknown sources possibly situated in the spiral arms, such as the two at the tail of the Sagittarius-Carina arm in the fourth quadrant, we believe that this does not impact the overall categorization of Unknown sources to the inter-arm regions.

\subsection{Detection and physical properties of ammonia gas} \label{subsec:3.2}
\begin{figure}[htbp!]
    \centering
    \includegraphics[width=2.0in]{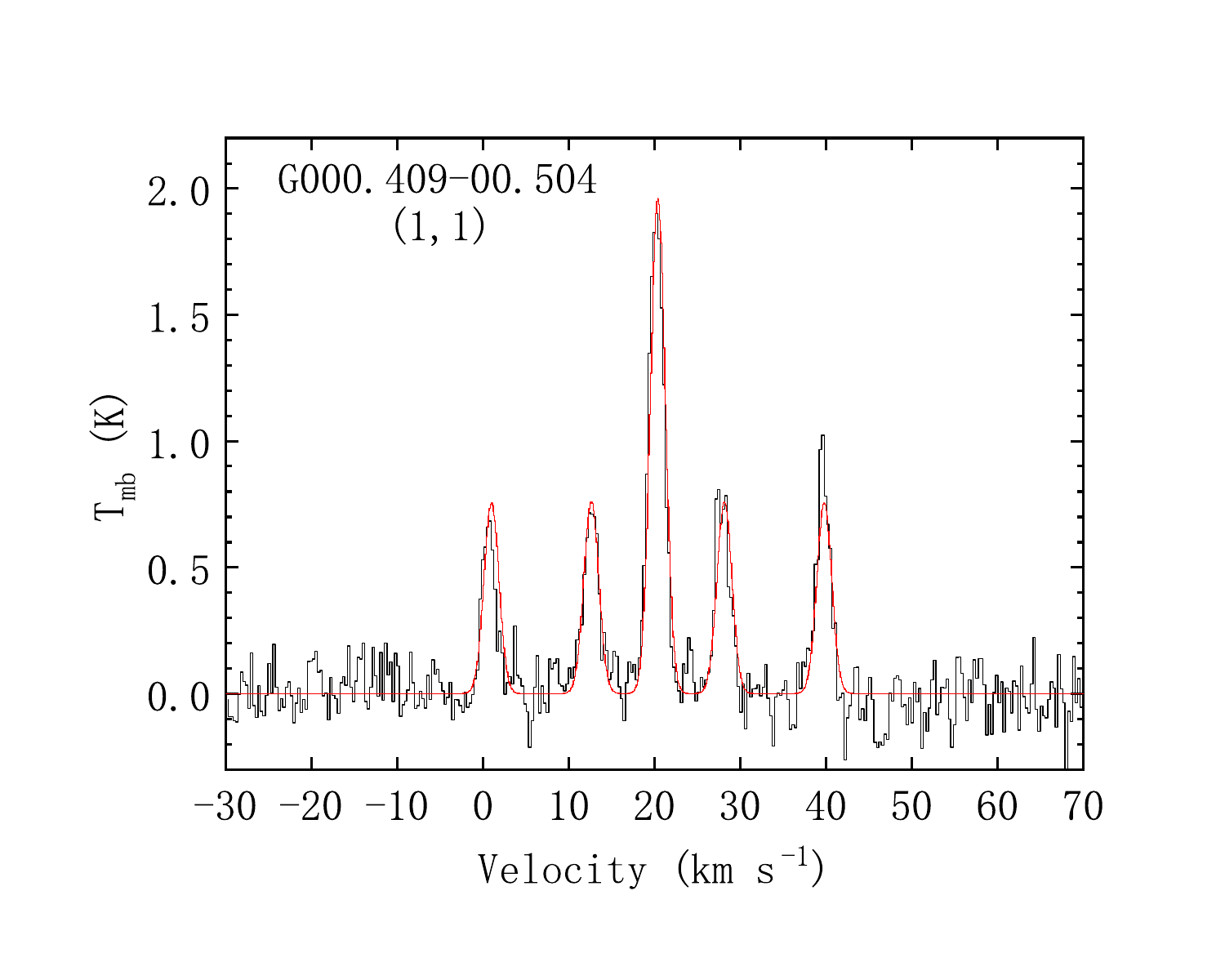}
    \includegraphics[width=2.0in]{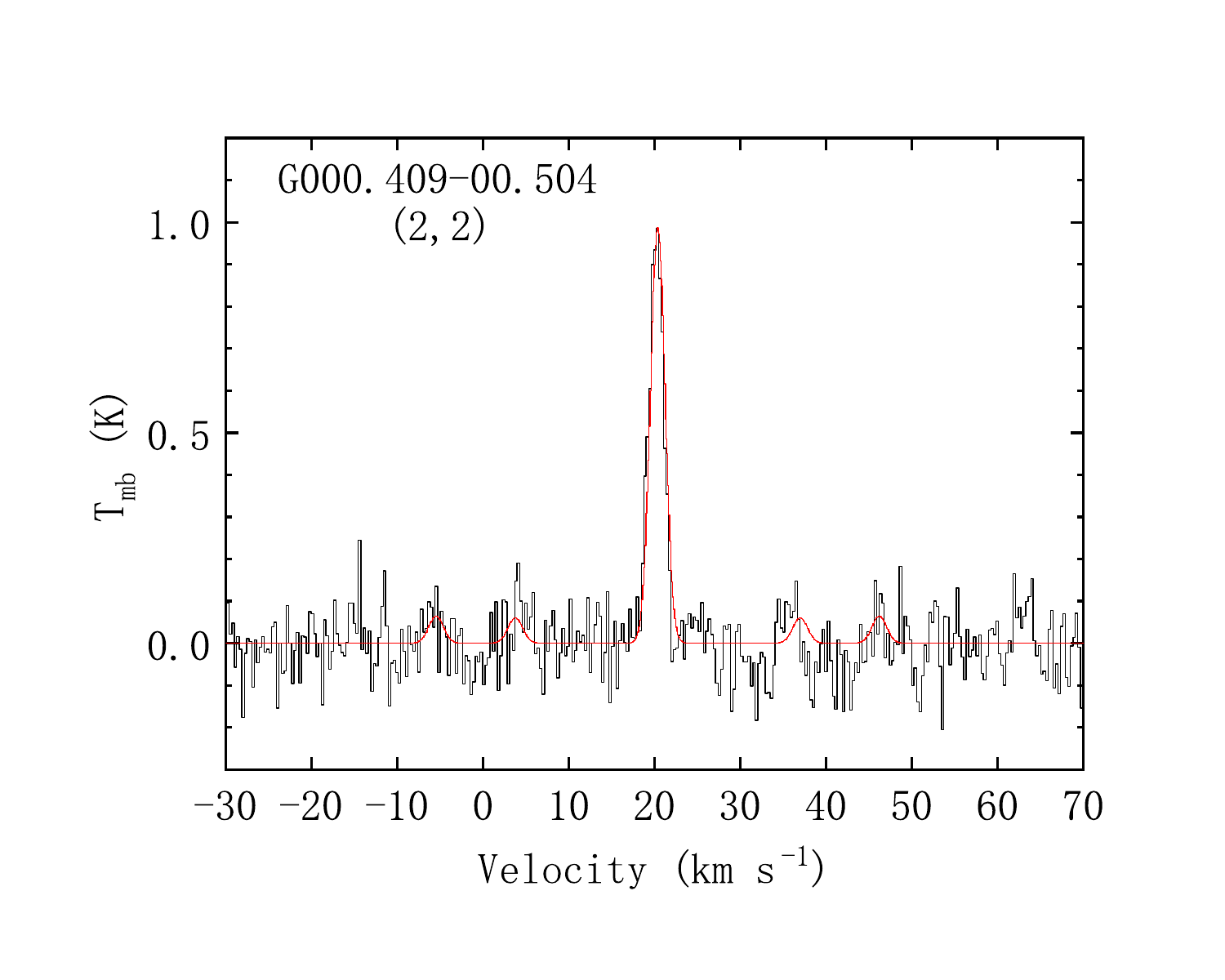}
    \includegraphics[width=2.0in]{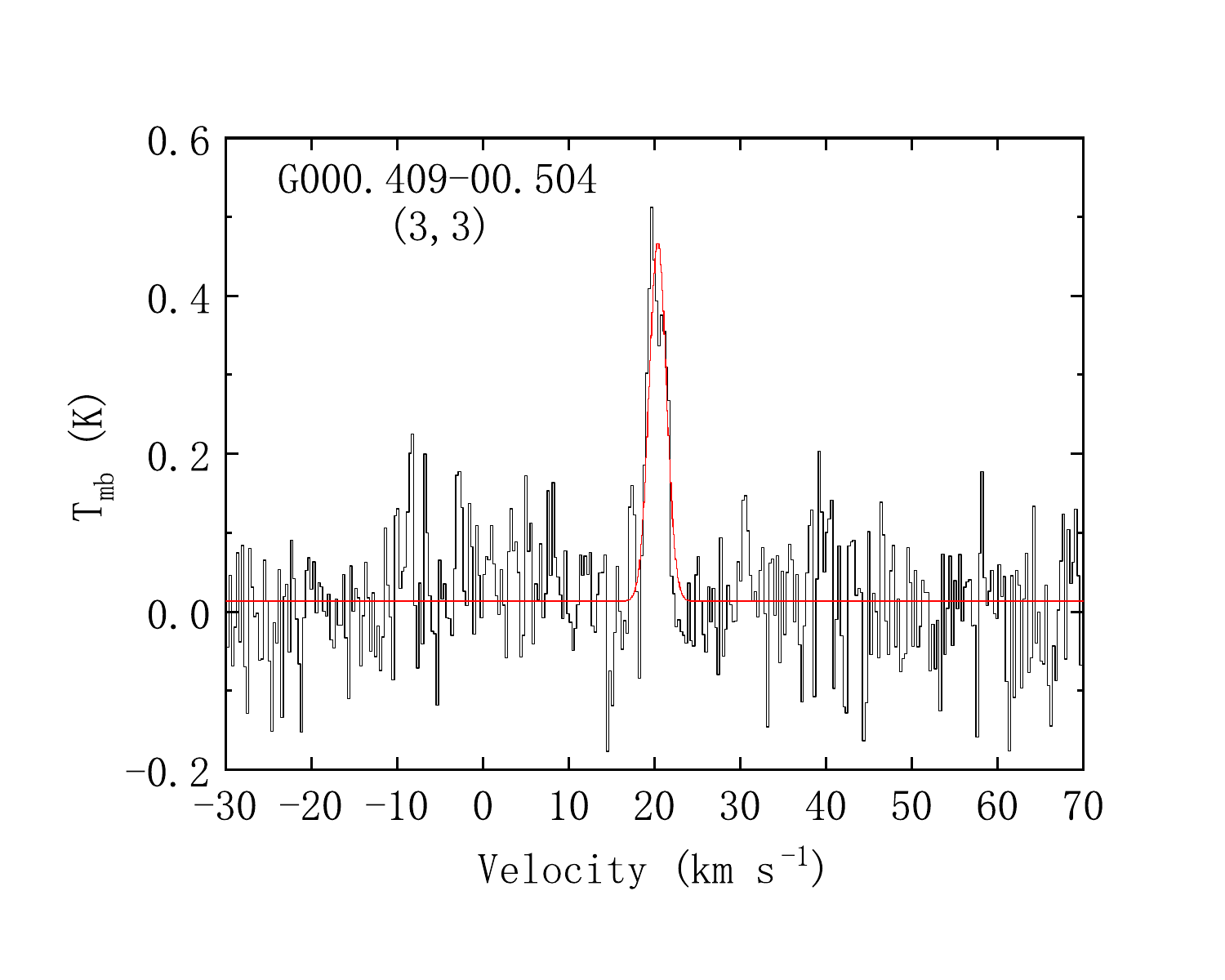}
    \caption{Spectra of NH$_3$ (1, 1), NH$_3$ (2, 2), and NH$_3$ (3, 3) were detected by TMRT. The red profiles represent the Gaussian fitting to the ammonia spectra. An example for G000.409$-$00.504 is presented, the extended version of this figure is available in the online journal.}
    \label{fig Spectra of nh3}
\end{figure}

Out of the observed 214 sources, 169, 158, and 126 were detected with the NH$_3$ (1, 1), NH$_3$ (2, 2), and NH$_3$ (3, 3) transitions, corresponding to the detection rates of 79.0\%, 73.8\%, and 58.9\%, respectively. Figure \ref{fig Spectra of nh3} displays the detected spectra of NH$_3$ (1, 1), NH$_3$ (2, 2), and NH$_3$ (3, 3). Out of the 169 sources of ammonia detected, 101 are also associated with 6.7 GHz methanol masers. 

The ammonia molecule is a reliable tracer for physical environments of a molecular cloud core \citep{2013ApJS..206...22C,2020ApJ...898..157M}. By observing the ammonia molecule, it is able to derive the physical parameters of the molecular cloud, such as excitation temperature and column density. The derived physical parameters from the detected ammonia transitions towards our observed sample can be found in Table \ref{table3}. The methods for deriving parameters of ammonia optical depth, temperature, and column density are briefly described as follows.

\begin{deluxetable*}{ccccrrrrrrc}
\tablecaption{Observations of ammonia molecules\label{table3}}
\tablehead{
\colhead{Name} & \colhead{Transition} & \colhead{Integration time} & \colhead{Rms} & \colhead{Area} & \colhead{$V_{p}$} & \colhead{$\Delta$V} & \colhead{S$_{p}$} & \colhead{$\tau$} & \colhead{T$_{ex}$} & \colhead{N} \\
\colhead{($^{\circ}$ $^{\circ}$)} & \colhead{} & \colhead{(min)} & \colhead{(mK)} & \colhead{(K km s$^{-1}$)} & \colhead{(km s$^{-1}$)} & \colhead{(km s$^{-1}$)} & \colhead{(K)} & \colhead{} & \colhead{(K)} & \colhead{(10$^{13}$ cm$^{-2}$)} \\
}
\decimalcolnumbers
\startdata    
G000.376$+$00.040 & NH$_3$ (1, 1) & 10.5  & 106 & 28.9$\pm$3.5  & 35.00$\pm$0.11    & 14.60$\pm$0.43 & 1.88  & 0.91$\pm$0.09 & 30.87 & 151.2 \\
                  & NH$_3$ (2, 2) & 10.5  & 88  & 19.1$\pm$0.5  & 35.30$\pm$0.13    & 11.50$\pm$0.25 & 1.42  &               &       &       \\
                  & NH$_3$ (3, 3) & 10.5  & 86  & 28.6$\pm$0.5  & 35.20$\pm$0.08    & 11.90$\pm$0.19 & 1.31  &               &       &       \\
G000.409$-$00.504 & NH$_3$ (1, 1) & 10.5  & 98  & 4.3$\pm$0.1   & 20.30$\pm$0.02    & 1.69$\pm$0.05  & 1.94  & 1.76$\pm$0.18 & 17.61 & 38.0  \\
                  & NH$_3$ (2, 2) & 10.5  & 74  & 2.1$\pm$0.1   & 20.30$\pm$0.04    & 1.84$\pm$0.14  & 0.99  &               &       &       \\
                  & NH$_3$ (3, 3) & 10.5  & 67  & 1.2$\pm$0.1   & 20.40$\pm$0.09    & 1.63$\pm$0.17  & 0.46  &               &       &       \\
G000.527$+$00.182 & NH$_3$ (1, 1) & 7.5   & 180 & 5.6$\pm$0.2   & $-$3.37$\pm$0.12  & 5.30$\pm$0.20  & 0.82  & 1.85$\pm$0.26 & 23.81 & 39.6  \\
                  & NH$_3$ (2, 2) & 7.5   & 101 & 3.3$\pm$0.2   & $-$3.64$\pm$0.15  & 4.75$\pm$0.33  & 0.61  &               &       &       \\
                  & NH$_3$ (3, 3) & 7.5   & 92  & 2.7$\pm$0.2   & $-$2.71$\pm$0.27  & 7.16$\pm$0.63  & 0.36  &               &       &       \\
\enddata
\tablecomments{Column(1): source name; Column(2): ammonia transitions; Column (3): total integration time; Column (4): the rms noise; Column (5): integration intensity; Columns (6) -- (8): the peak velocity, line width (FWHM), and main-beam temperature, respectively; Columns (9) -- (11): the derived optical depth, excitation temperature and column density of ammonia.\\
(This table is available in its entirety in machine-readable form.)}
\end{deluxetable*}

\begin{itemize}
    \item \textbf{Optical depth:} The fine structure lines of the transitions of the NH$_3$ metastable levels, such as NH$_3$ (1, 1), NH$_3$ (2, 2), and NH$_3$ (3, 3), are easily detected in the  molecular cloud cores compared to other transitions. The relative intensity between the fine structure lines remains unchanged initially, but it undergoes a change after passing through the molecular cloud due to optical depth effect. By comparing the changes in the relative intensity of the fine structure lines, it is possible to obtain optical depth, $\tau$:
    \begin{equation}
        \frac{T_B(J,K,m)}{T_B(J,K,s)} = \frac{1-e^{-{\tau}_{\nu}(J,K,m)}}{1-e^{-\alpha{\tau}_{\nu}(J,K,m)}},
    \end{equation}
    where $T_B(J, K, m)$ and $T_B(J, K, s)$ represent the peak brightness temperature of the main line and the satellite line, respectively, $\tau_{\nu}(J, K, m)$ represents the optical depth of the main line group, and $\alpha$ is the intensity ratio of the satellite line and main line. 
    
    \item \textbf{Temperature:} 
    After obtaining the optical depth $\tau$(1, 1, m) of the main line group of NH$_3$ (1, 1), the rotational temperature can be calculated using the following formula provided by \citet{2018A&A...616A.111W}:
    \begin{equation}
        T_{rot}(2,2:1,1) = \frac{-41.5}{ln[\frac{-0.282}{\tau(1,1,m)}ln(1-\frac{T_{B}(2,2,m)}{T_{B}(1,1,m)})(1-e^{-\tau(1,1,m)})]},
    \end{equation}
    where $T_{B}(1, 1, m)$ and $T_{B}(2, 2, m)$ represent the peak brightness temperatures of the main lines at (1, 1) and (2, 2) transitions, respectively. In the local thermodynamic equilibrium (LTE) state, the excitation temperature is generally equivalent to the rotational temperature \citep{2011MNRAS.418.1689U}.
    
    \item \textbf{Column density:} In the LTE condition, the column density can be calculated using the following formula \citep{2013ApJS..206...22C}:
    \begin{equation}
        N = \frac{3kW}{8 \pi ^3 \nu S \mu ^2}(\frac{T_{ex}}{T_{ex}-T_{bg}})(\frac{\tau}{1-e^{-\tau}})Q(T_{ex})e^{\frac{E_{u}/k}{T_{ex}}},
    \end{equation}
    where $k$ is Boltzmann's constant in erg K$^{-1}$, $W$ represents the integrated intensity of the observed spectral line in units of K km s$^{-1}$, $\tau$ is the optical depth. $\nu$ is the frequency of the observed spectral line in Hz,  $S\mu^2$ is the product of line strength and the square of the electric dipole moment, $E_u/K$ represents the upper energy level of NH$_3$ (1, 1) in K. These three values can be found on SPLATALOGUE\footnote{\url{https://splatalogue.online//advanced.php}}. $T_{bg}$ is the background brightness temperature, with a value of 2.73 K. $Q(T_{ex})$ represents the partition function at the excitation temperature. We adopted partition function $Q(T_{ex})$ = 0.1261 $\times$ $T^{1.478}_{ex}$ in our analysis.
\end{itemize}

Of the 158 sources with  NH$_3$ (2, 2) transition detected, two sources (G043.166$+$00.011, G209.007$-$19.385) were excluded in the calculation because the emission of the NH$_3$ (1, 1) transition was lower than that of (2, 2) transition in the two sources. Therefore, a total of 156 sources were derived the physical properties. The derived optical depths in our sample are in a range of 0.10 to 6.37, with a mean of 1.63 $\pm$ 1.09. The excitation temperatures are derived in a range of 8.0 K to 88.5 K, with a mean of 20.1 $\pm$ 8.8 K. The NH$_{3}$ column densities span from $2.5 \times 10^{13}$ cm$^{-2}$ to $7.0 \times 10^{15}$ cm$^{-2}$, with a mean value of $(7.3 \pm 8.6) \times 10^{14}$ cm$^{-2}$.

\section{Discussions} \label{sec:4}
\subsection{Large-scale distribution characteristics of 6.7 GHz methanol masers in the Milky Way} \label{subsec:4.1}
\subsubsection{Observation effects of the 6.7 GHz methanol masers} \label{subsubsec:4.1.1}
In the surveys, the telescopes were only able to detect distant sources with high luminosity, while those with low luminosity would be missed once their luminosities are lower than the survey sensitivity limit. 
Therefore, there is a general observational effect in the surveys. In order to explore the impact of such the observation effect on our newly collected 6.7 GHz methanol maser sample, we analyzed the distributions of heliocentric distance and spiral angle for the maser sources.

\begin{figure}[htbp!]
    \centering
    \includegraphics[width=0.5\linewidth]{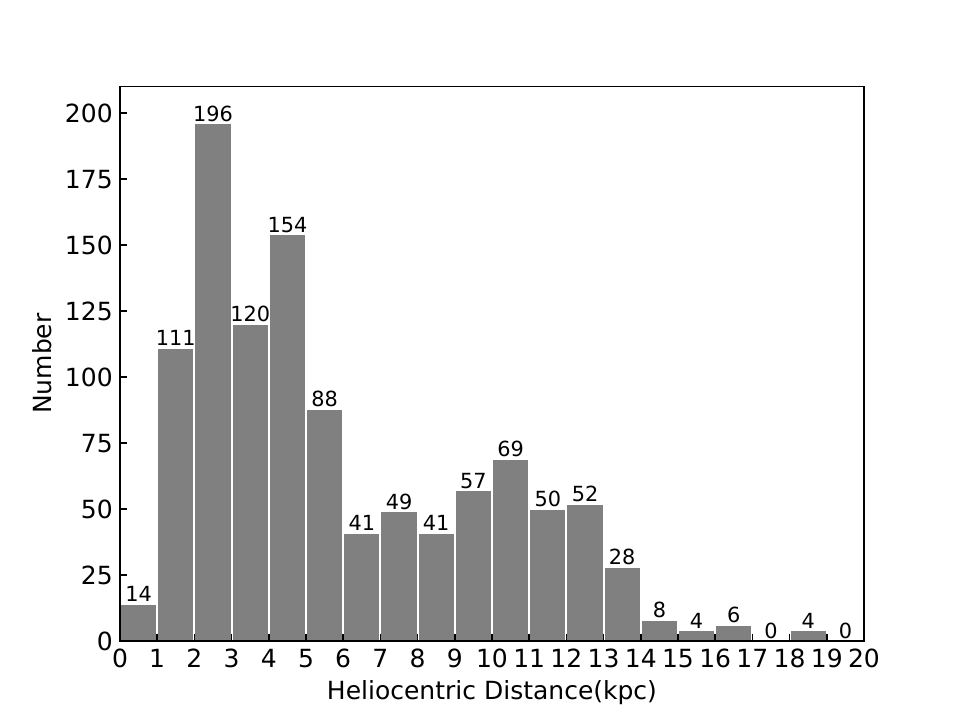}
    \caption{Histogram distributions of the heliocentric distances of the 6.7 GHz methanol masers in the Milky Way.}
    \label{fig D_sun}
\end{figure}

In Figure \ref{fig D_sun}, we present histogram showing the distribution of heliocentric distances of 6.7 GHz methanol masers. The most abundant masers are found at distances between 1 and 6 kpc from the solar system, indicating that they are most dense in the vicinity of the solar system. As the heliocentric distance increases, the observed region passes through the center of the Galaxy, reaching the other side of the Galaxy, and the number of observed masers decreases by approximately half at distances of 6--13 kpc. 
The histogram distribution of methanol maser number along the heliocentric distance is generally agrees with that of the dust clumps derived from \cite{2019MNRAS.490.2779B}. This suggests that the number of observed 6.7 GHz methanol masers progressively decreases as the observation distances. While it is possible that the 6.7 GHz methanol masers are sparsely distributed on the far side of our Galaxy, with fewer in number than on the near side, it is more likely that the above observation effect is responsible for this decrease in the number of observed masers.

\begin{figure}[htbp!]
    \centering
    \includegraphics[width=0.5\linewidth]{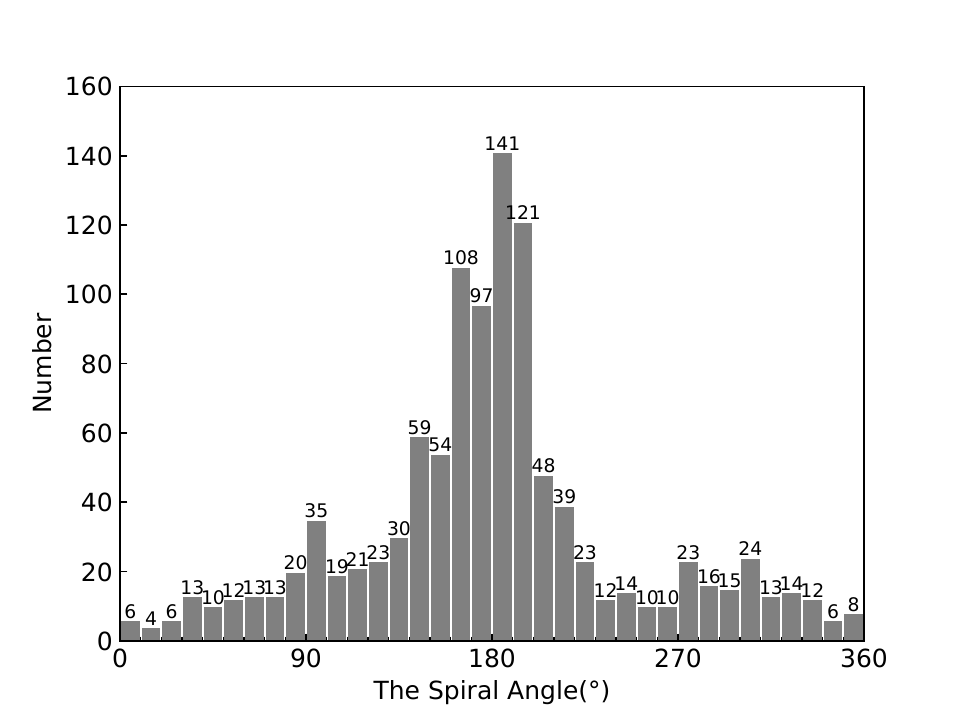}
    \caption{Distribution of the spiral angles of the 6.7 GHz methanol masers in the Milky Way. Centered on the galactic center, the spiral angle is defined with 0$^{\circ}$ set as the direction opposite to the Sun, increasing counterclockwise, with the Sun located at 180$^{\circ}$.}
    \label{fig Spiral angle}
\end{figure}

Figure \ref{fig Spiral angle} shows the distribution of the 6.7 GHz methanol masers in various directions around the Galactic center, with the solar system located at the 180$^{\circ}$ direction (see Figure \ref{fig Overlaying}). It is apparent that the distribution of 6.7 GHz methanol masers is highly concentrated in the angle range of 160$^{\circ}$--200$^{\circ}$, indicating that the current catalog of 6.7 GHz methanol maser sources is non-isotropic and highly concentrated towards the direction of the solar system. Assuming that 6.7 GHz methanol masers are uniformly distributed throughout the Galaxy, it is evident that there is a significant observation effect in the currently compiled catalog of 6.7 GHz methanol maser sources. It can be believed that there are still many undetected 6.7 GHz methanol masers in the direction intervals of 0$^{\circ}$--160$^{\circ}$ and 200$^{\circ}$--360$^{\circ}$.

To minimize the impact of the above observation effect in subsequent analysis, it is necessary to exclude sources with large distances from the 6.7 GHz methanol maser catalog. Given that the distance between the solar system and the galactic center is 8.15 kpc, sources on the opposite side of the Milky Way are thought to be located at distances greater than 8.15 kpc. Furthermore, the number of sources located within the spiral angle range of $-90^{\circ}$ to $90^{\circ}$ is very small, indicating that these sources are particularly susceptible to the observation effect. Thus, we excluded these sources and used the remaining sources with spiral angles ranging from 90$^{\circ}$ to 270$^{\circ}$ as our sample for subsequent analysis. This reduced sample consists of 864 sources.

\subsubsection{Distribution of 6.7 GHz methanol masers along the galactocentric distance} \label{subsubsec:4.1.2}
So far, the source with the largest galactocentric distance in the 6.7 GHz methanol maser catalog is G168.060 $+$ 00.820. It has a galactocentric distance of 13.43 kpc and is located at the tail of the Norma-Outer arm. Therefore, all 6.7 GHz methanol masers in the catalog are located within a distance of 14 kpc from the Galactic center.

\begin{figure}[htbp!]
    \centering
    \begin{tabular}{@{}cc@{}}
    \includegraphics[width=0.5\linewidth]{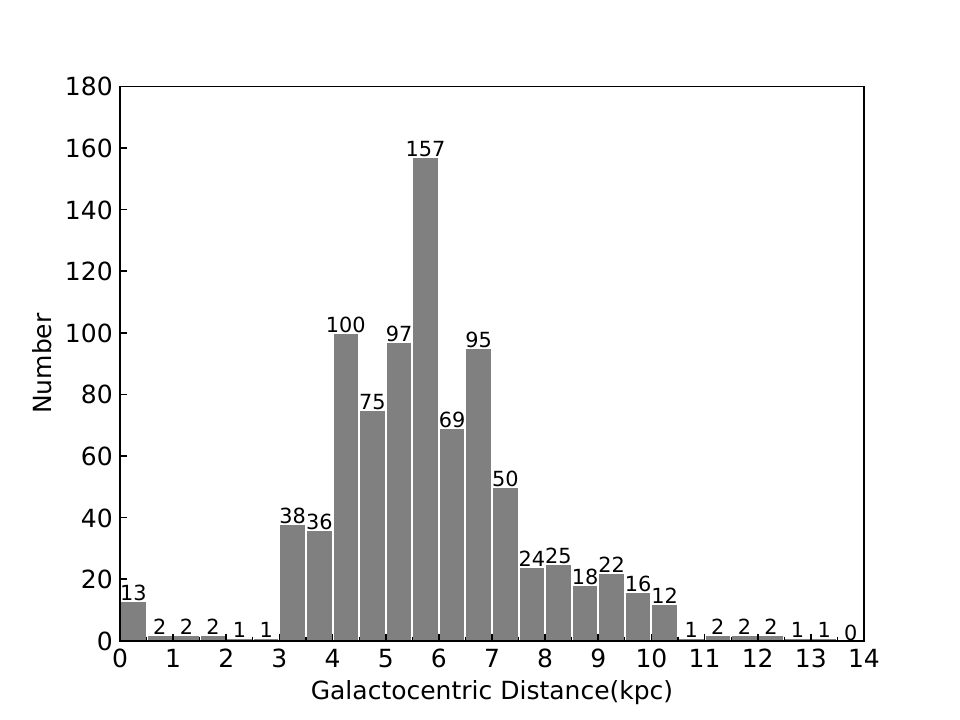}\\
    \end{tabular}
    \caption{Histogram distribution of the galactocentric distances of the reduced 6.7 GHz methanol maser sample in the Milky Way.}
    \label{fig D_GC}
\end{figure}

Figure \ref{fig D_GC} displays histogram distributions of the galactocentric distances of 6.7 GHz methanol maser sources after correcting for the observation effect. The distribution shows a concentration of sources in the range of 3.0--10.5 kpc, which is consistent with the consensus that the spiral arm structure of the Galaxy exists at distances between approximately 3 kpc and 10 kpc from the Galactic center \citep{2021A&A...645L...8X}. Within the range between 4 kpc and 7 kpc, the distribution density of 6.7 GHz methanol masers is most pronounced, indicating the highest level of high-mass star formation activity and a greater abundance of material to form high-mass stars in these regions. We have also found a few 6.7 GHz methanol maser sources in the regions close to the Galactic center within 3 kpc. As the assumption of flat rotation curve fails in the inner part of the Galaxy, the distances of sources with galactocentric distances less than 3 kpc have large uncertainties. To obtain accurate distances, more precise astrometric measurements are required. As far as the current distribution of methanol masers is concerned, the distribution density of methanol masers in the inner Galaxy is considerably lower compared to those in the spiral arm regions. This suggests that the ongoing high-mass star formation activity is suppressed in the central region of the Milky Way compared to the spiral arms, which is supported by the other findings. \citet{2013MNRAS.431.1752U} support that star-forming efficiency is significantly lower in the Galactic Centre region compared to the rest of their surveyed area. \citet{2013MNRAS.429..987L} found that the ratio of star formation indicators to the amount of dense gas is lower in the central molecular zone (CMZ) compared to the Galactic disc, suggesting that star formation activity in the CMZ is suppressed or relatively less active than that in the Galactic disc. Additionally, \citet{2019MNRAS.482.5349R} also found the density of 6.7 GHz methanol masers in the CMZ is relatively low compared to other regions, and they suggested that this could be due to the unique physical properties of the CMZ, such as higher molecular gas density, higher temperature, and increased turbulent velocity. These properties may inhibit star formation in the CMZ, resulting in a lower density of star-forming sites. Beyond 10.5 kpc, the 6.7 GHz methanol masers are scarce, and these areas correspond to the tails of the main spiral arms, indicating a significant reduction in the formation of high-mass stars therein.

\subsection{Characteristics of the 6.7 GHz methanol maser luminosity} \label{subsec:4.2}
\subsubsection{Luminosity function of the 6.7 GHz methanol masers} \label{subsubsec:4.2.1}
\begin{figure}[htbp!]
    \centering
    \includegraphics[width=0.5\linewidth]{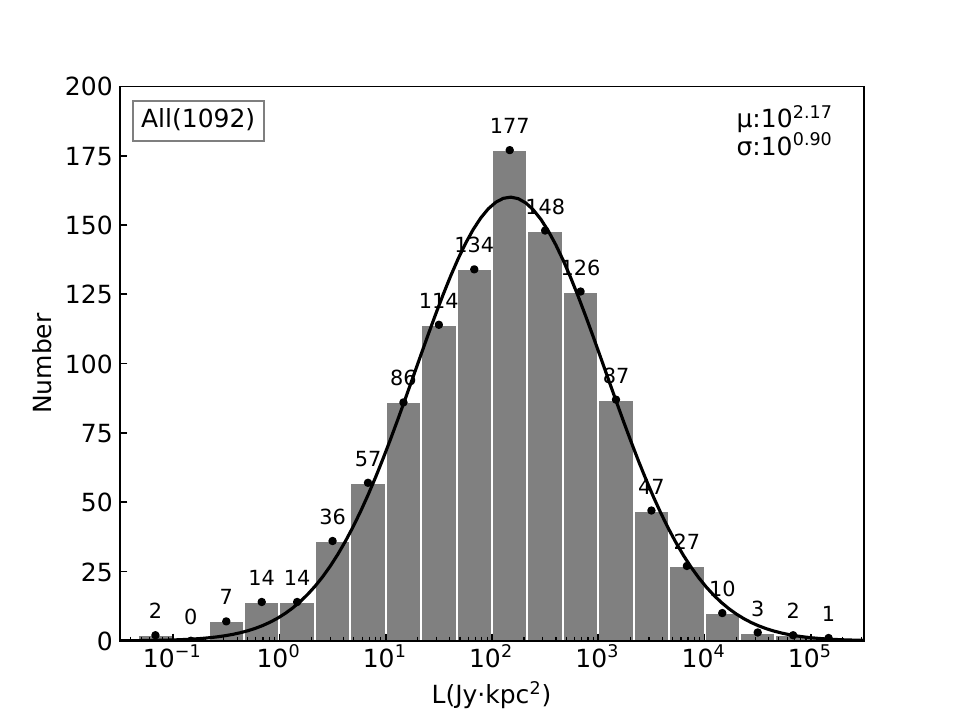}
    \caption{Histogram of the luminosity of all the 1092 6.7-GHz methanol masers in the Milky Way. The solid line represents the fitting line for their log-normal distribution. The legend on the upper right displays the mean value ($\mu$) and standard deviation ($\sigma$) of the distribution.}
    \label{fig L}
\end{figure}

The investigation of the luminosity function of 6.7 GHz methanol masers in the Milky Way can provide important reference for further exploring methanol masers in extragalactic galaxies. Figure \ref{fig L} displays a luminosity histogram containing all the collected 1092 methanol masers in the Milky Way. After conducting a K-S test, we observed the log-luminosity distribution adhering to a normal pattern, implying a certain level of completeness in the existing 6.7 GHz methanol maser sample in the Milky Way from the view of luminosity distribution. Therefore, this also indicates that the undetected methanol masers (see Section \ref{subsubsec:4.1.1}) should not be very bright.

\subsubsection{The relationship between 6.7 GHz methanol maser luminosity and spiral angle} \label{subsubsec:4.2.2}
\begin{figure}[htbp!]
    \centering
    \begin{tabular}{@{}cc@{}}
    \includegraphics[width=0.4\linewidth]{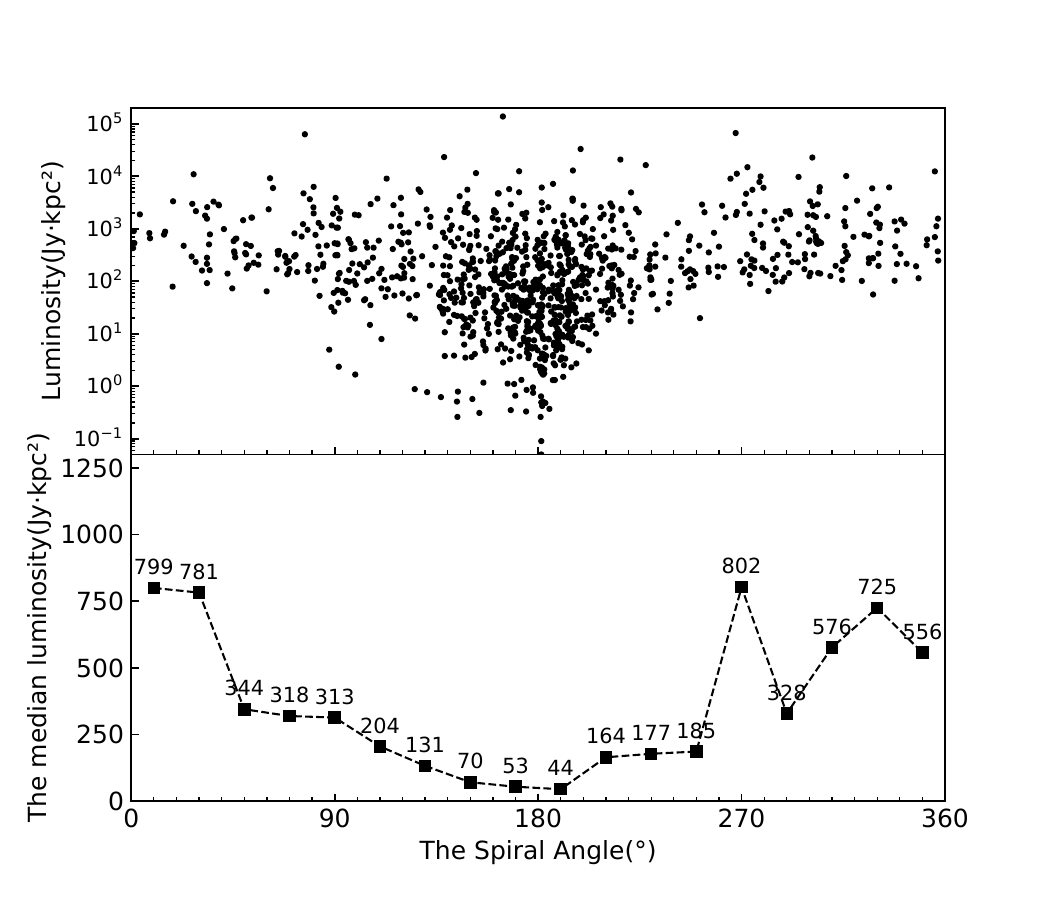}
    \includegraphics[width=0.4\linewidth]{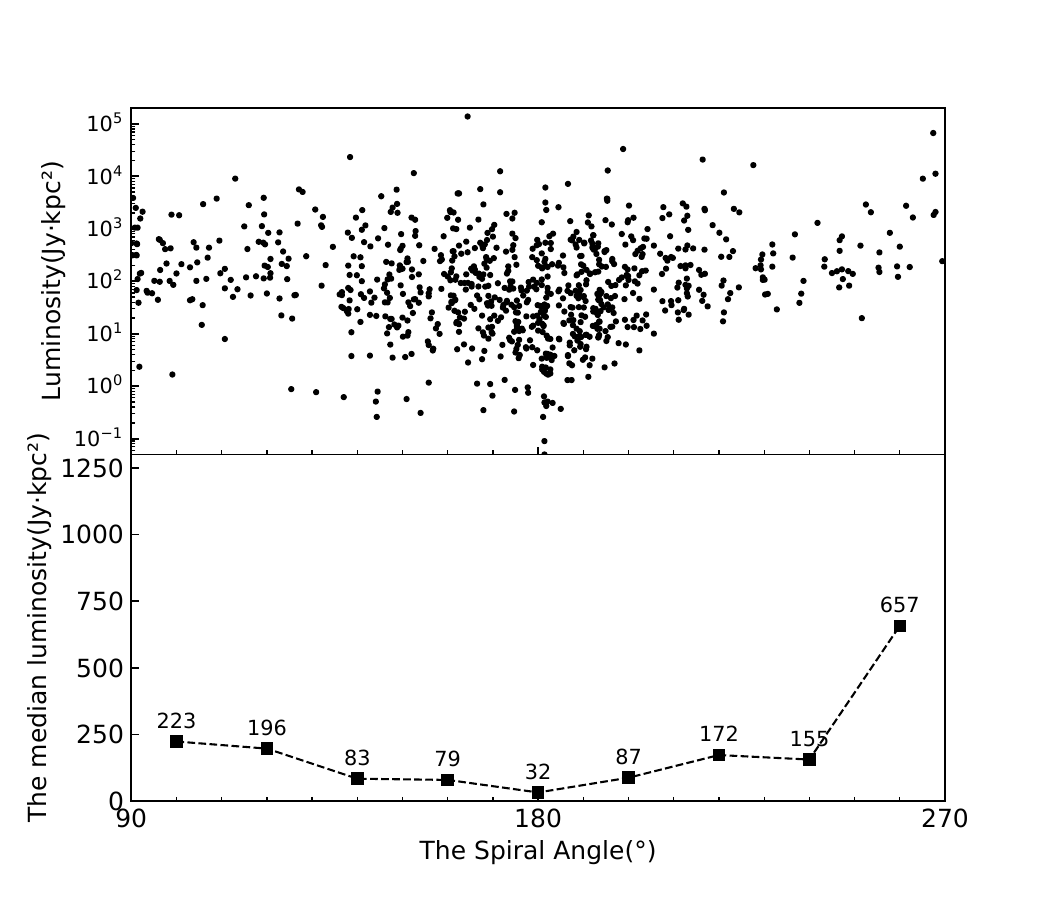}\\
    \includegraphics[width=0.4\linewidth]{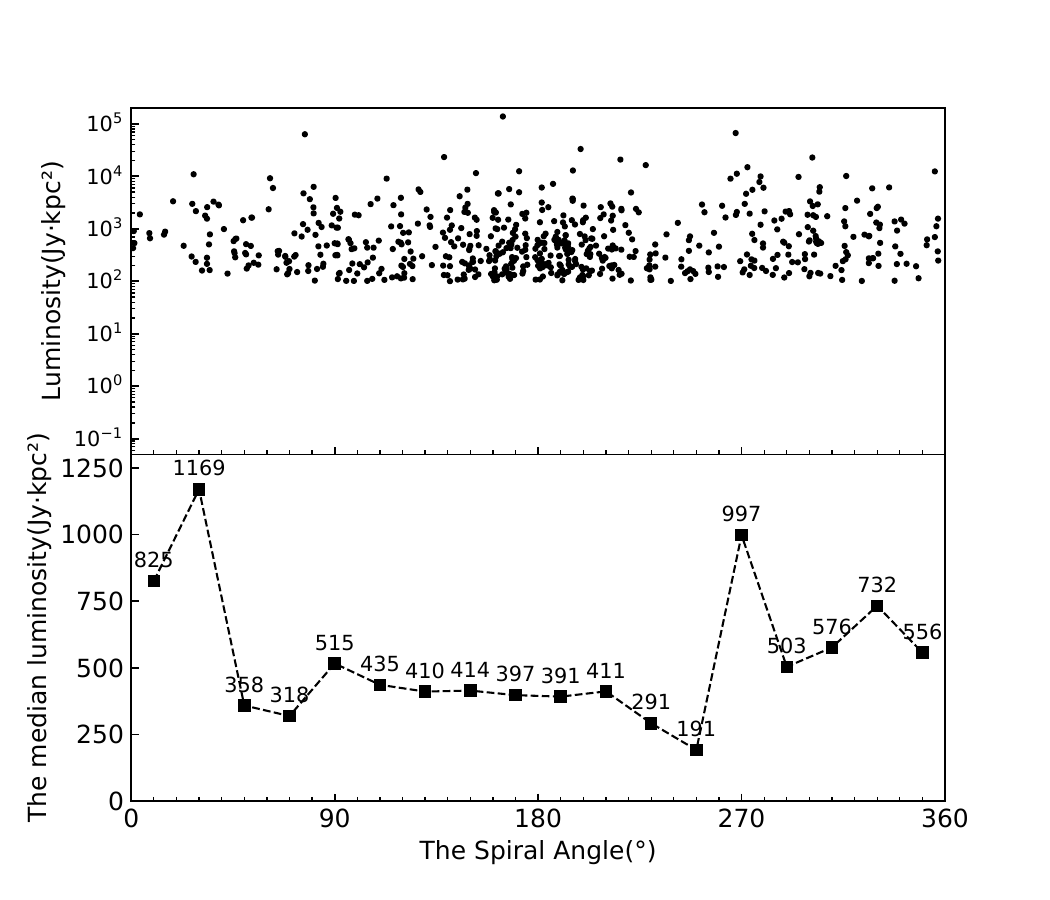}
    \includegraphics[width=0.4\linewidth]{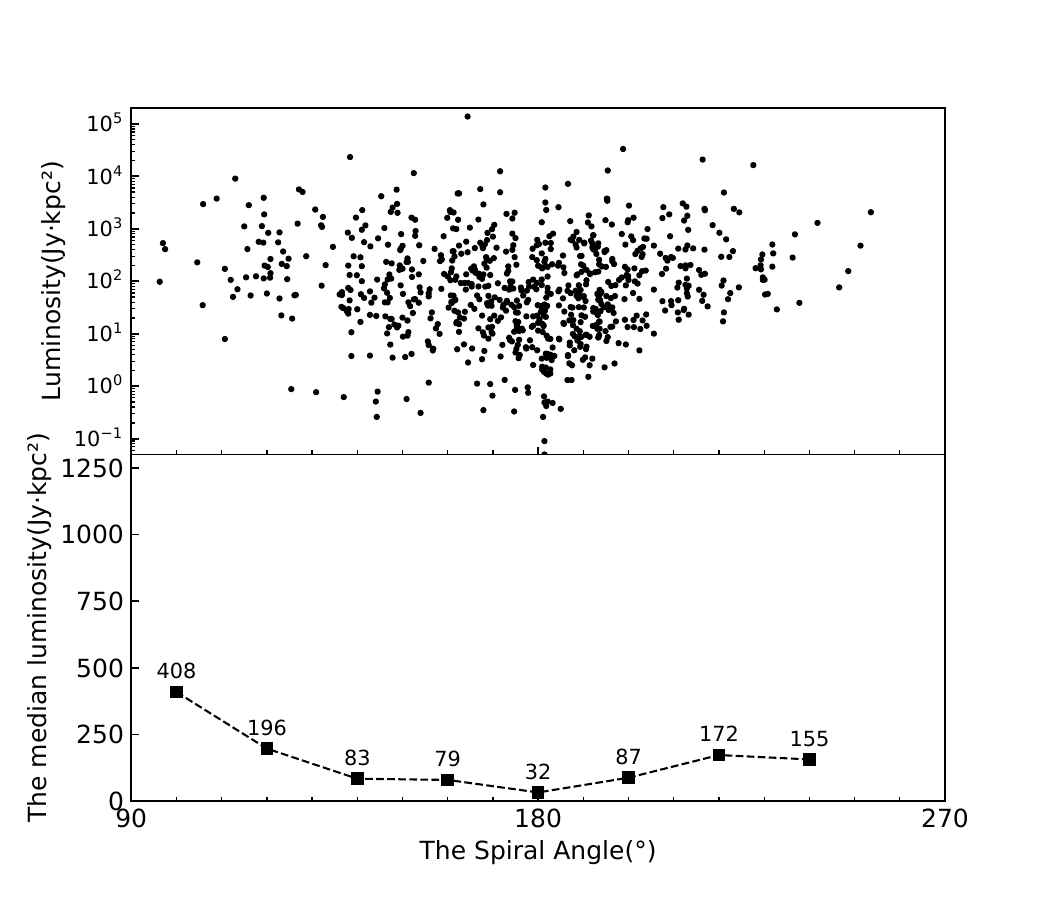}
    \end{tabular}
    \caption{Luminosity of 6.7 GHz methanol masers as a function of spiral angle. Top-left panel: 1092 sources, without correction for the observation effect. Top-right panel: 864 sources, after correcting for the observation effect of retaining sources with SA = 90$^{\circ}$--270$^{\circ}$. Bottom-left panel: 628 sources, after correcting for the observation effect of retaining sources with luminosity $ge$ 100 Jy$\cdot$kpc$^2$. Bottom-right panel: 784 sources, after correcting for the observation effect of retaining sources with heliocentric distance $<$ 8.15 kpc.} The top half of the panels consists of scatter plots showing the relationship between the 6.7 GHz methanol maser luminosity and spiral angle. The bottom half features a line plot illustrating the median luminosity along the spiral angle, with data points averaged at an intervals of 20$^{\circ}$.
    \label{fig All arm}
\end{figure}

Figure \ref{fig All arm} was generated to analyze the luminosity distribution of the 6.7 GHz methanol masers in all directions surrounding the Galactic center. As shown in the top-left panel of Figure \ref{fig All arm}, the distribution of methanol masers is denser at the spiral angle of 180$^{\circ}$, which corresponds to the direction of the solar system, with the average luminosity of about 50 Jy$\cdot$kpc$^2$. In contrast, toward the regions with the spiral angles of 0$^{\circ}$--90$^{\circ}$ and 270$^{\circ}$--360$^{\circ}$, the detected methanol masers have a higher luminosity, but the number of masers is significantly lower. This further indicates that the current detection of 6.7 GHz methanol masers is incomplete. There are still numerous low-luminosity methanol masers that have not been discovered in the direction opposite to the solar system. After correcting for the observation effect (see Section \ref{subsubsec:4.1.1}), the top-right panel of Figure \ref{fig All arm} reveals that the median luminosity from the spiral angle of 90$^{\circ}$--270$^{\circ}$ is nearly identical. This indicates that the simplified sample can effectively reduce the observation effect. 

In addition, we tried two other methods to correct the observed effect.  The first method is to eliminate sources with luminosity values below 100 Jy$\cdot$kpc$^2$, so that the maser luminosity range is identical in the region of spiral angles 0--360° as shown in the bottom-left panel of Figure \ref{fig All arm}, leaving 628 sources in the catalog. Considering that the results obtained for this simplified sample are only applicable to sources with luminosity values greater than 100 Jy$\cdot$kpc$^2$, and it is not possible to tell whether sources with lower luminosity values (i.e., regions of weaker star-formation activity) also obey these conclusions, we therefore did not adopt this approach. The second method is to retain only the sources within a heliocentric distance of 8.15 kpc, leaving 784 sources in the catalog as shown in the bottom-left panel of Figure \ref{fig All arm}. But the difference between the results obtained by this method and those obtained by the original method (which retains spiral angles from 90 to 270 degrees, leaving 864 sources; see Section 4.1.1) is not significant, and we adopted the original method.

\subsubsection{The luminosity distribution of 6.7 GHz methanol masers along galactocentric distance} \label{subsubsec:4.2.3}
\begin{figure}[th!]
    \centering
    \begin{tabular}{@{}cc@{}}
    \includegraphics[width=0.5\linewidth]{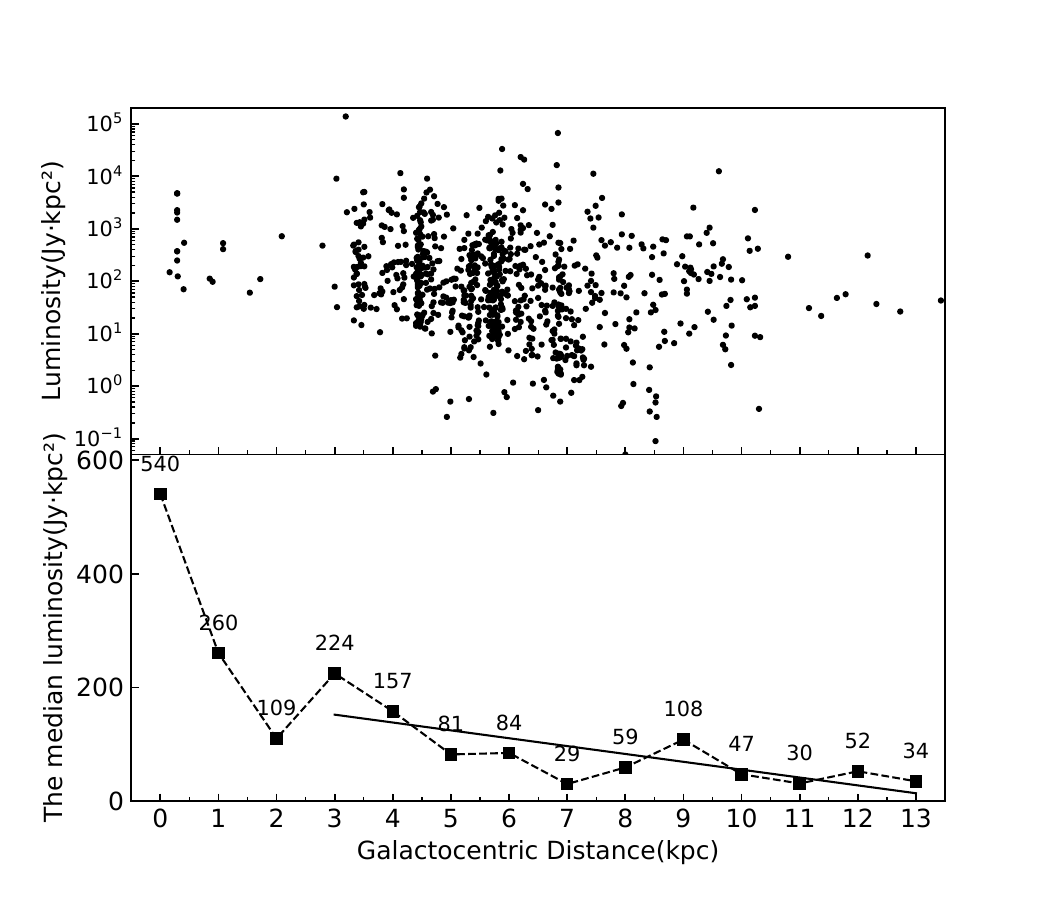}\\
    \end{tabular}
    \caption{Luminosity of 6.7 GHz methanol masers as a function of galactocentric distances. The scatter plot and the line plot of the median luminosity taken every 1 kpc are given in the upper and lower panels, respectively.}
    \label{fig L and D_GC}
\end{figure}

The luminosity of the 6.7 GHz methanol masers as a function of galactocentric distances is given in Figure \ref{fig L and D_GC}. Here, we used the reduced sample after correcting for the observation effect. From this figure, it can be clearly seen that the luminosity of 6.7 GHz methanol maser decreases as the galactocentric distance. Our finding indicates that the median luminosity of 6.7 GHz methanol masers shows a noticeable downward trend at distances up to 13 kpc from the Galactic center. This phenomenon suggests a higher level of activity of high-mass star formation toward the inner Milky Way, with activity decreasing as the galactocentric distances. The galactocentric distribution reveals that the surface density of the HMSFRs and the proportion of methanol-maser associated clumps are significantly higher in the inner Galaxy compared to the outer galaxy \citep{2013MNRAS.431.1752U,2015MNRAS.446.3461U}. Notably, despite the relatively scarce of 6.7 GHz methanol maser sources in the Galactic central region, their overall high luminosity suggests intense activities of high-mass star formation therein.

\subsubsection{Luminosity difference of 6.7 GHz methanol masers at four main spiral arms} \label{subsubsec:4.2.4}
\begin{figure}[b!]
    \centering
    \includegraphics[width=0.5\linewidth]{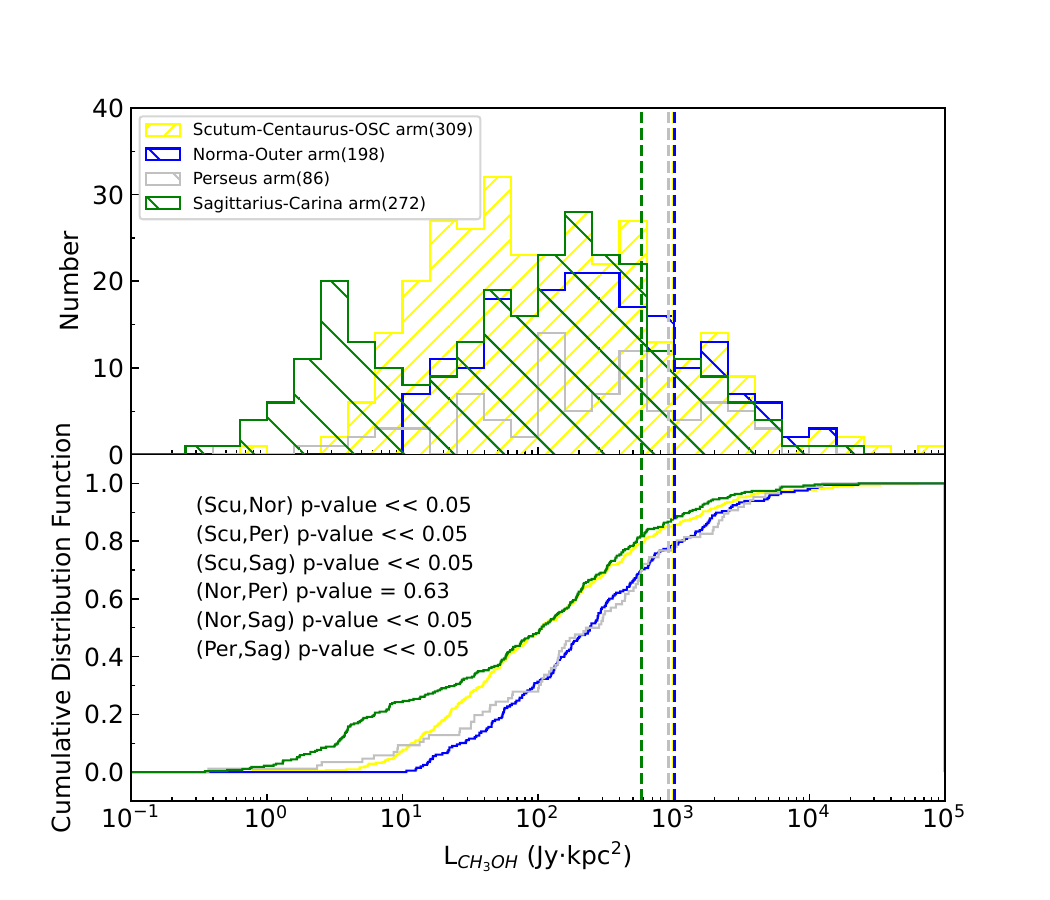}\\
    \includegraphics[width=0.4\linewidth]{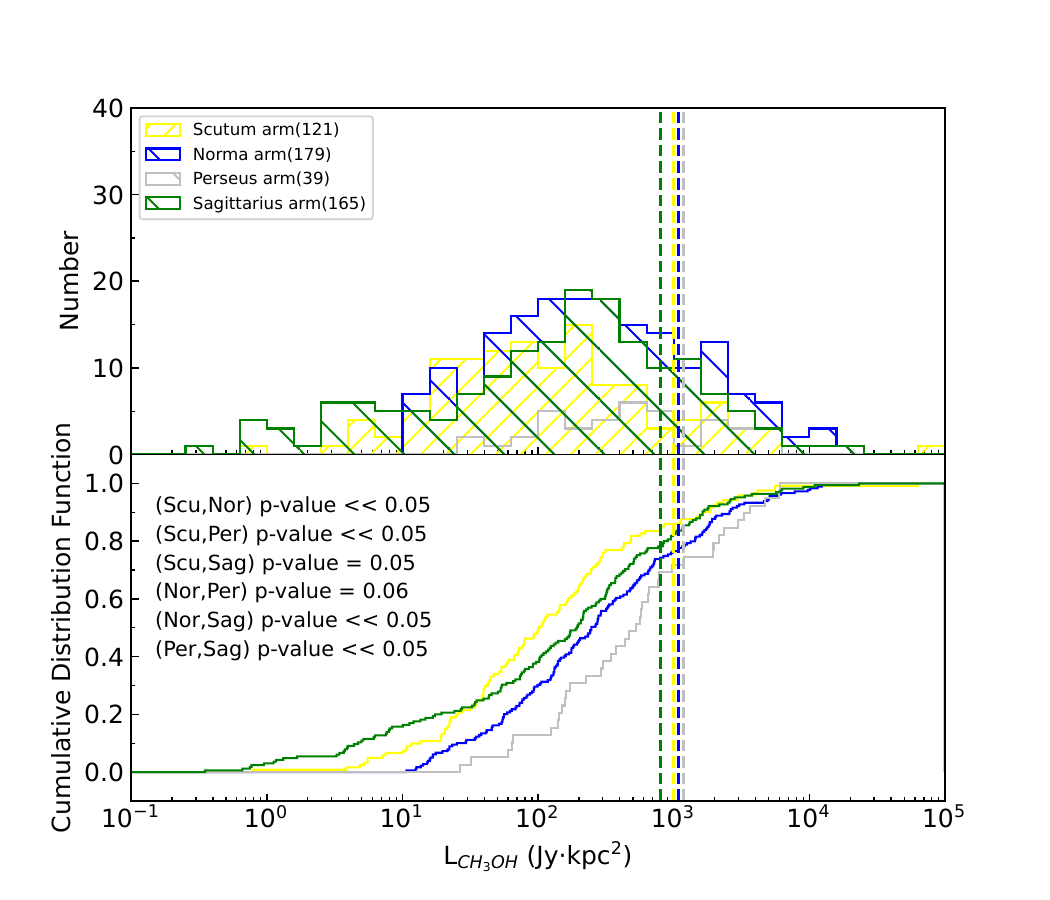}
    \includegraphics[width=0.4\linewidth]{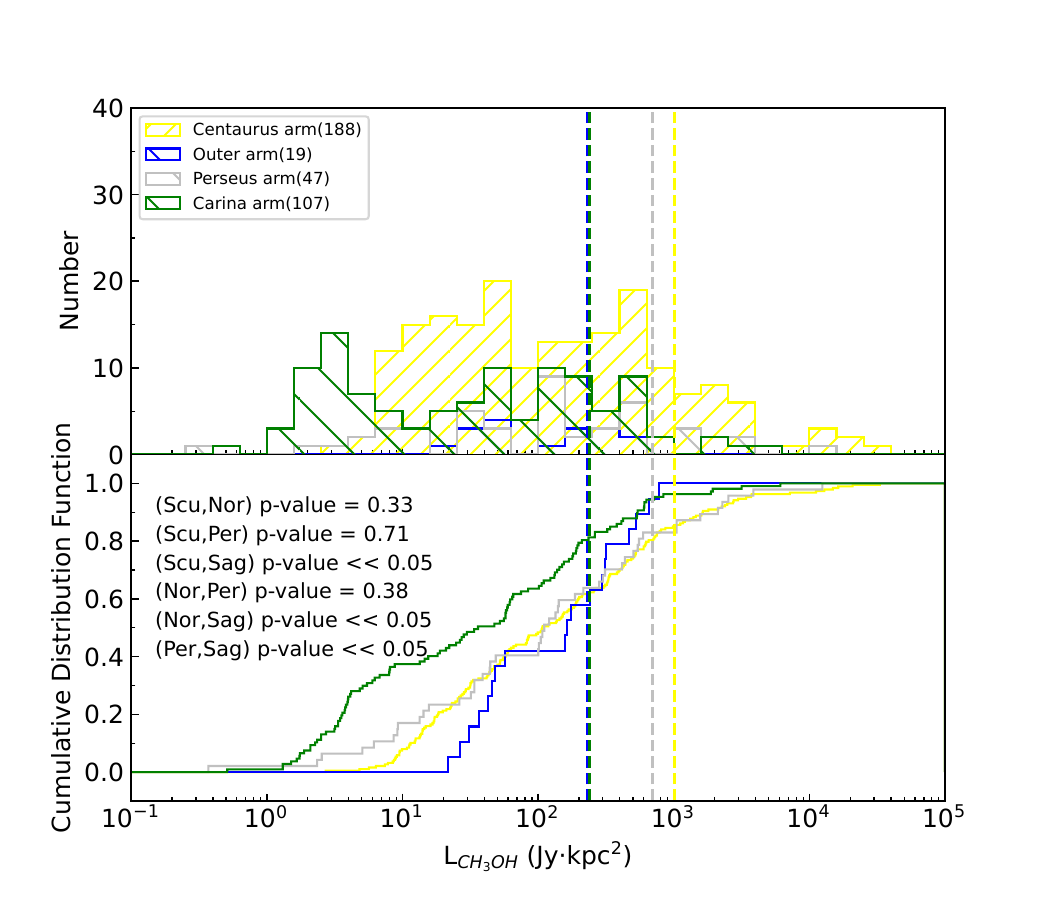}
    \caption{Comparison of the luminosity of the 6.7 GHz methanol masers at the four main spiral arms. The top panel shows four complete spiral arms, including inner and outer parts. The bottom-left panel shows the inner part of the four spiral arms, and the bottom-right panel shows the outer part of the four spiral arms. The legend in the upper left corner of the three panels describes the names of each arm and the number of sources located in each arm. The vertical dashed lines represent the median luminosity of masers at each arm. The p-values from the K-S tests are presented in each panel.}
    \label{fig L four arms}
\end{figure}

The latest BeSSeL project reveals that our Galaxy is composed of four main spiral arms. Therefore, it will be crucial to compare the characteristics of these four arms using masers or molecular lines. Considering the fact that, if a sample corrected for observational effects is used, the retained segments of the Perseus arm are the tail of the spiral arm, while the retained segments of the remaining three main spiral arms contain a large number of sources located at the head of the spiral arm, it may not be fair to compare them together. Therefore, we compared the luminosity differences between the four complete spiral arms, each of which contains sources from both the inner and outer parts. In addition, we also separately compared the luminosity differences of the inner and outer parts of the four spiral arms. According to the distance calculator's spiral arm classification (see Figure \ref{fig Overlaying}), the cut-off positions between the inner and outer arms are defined at: SA = 180$^{\circ}$ for the Scutum-Centaurus-OSC arm, SA = 360$^{\circ}$ for the Norma-Outer arm, and SA = 180$^{\circ}$ for the Sagittarius-Carina arm. Whereas, since the Perseus arm is not separated by default, we have set the Perseus arm's cut-off point as SA = 90$^{\circ}$. The initial catalog of 6.7 GHz methanol maser sources, uncorrected for the observational effect, was used for this analysis. As illustrated in Figure \ref{fig L four arms}, despite the comparatively lower quantity of methanol masers located in the Perseus arm compared to the other three main spiral arms, its luminosity is relatively higher. This result holds for the inner regions of the galaxy, the outer regions and the whole galaxy. This interesting phenomenon indicates the unique properties of the Perseus arm. The scarcity of masers from this arm may reflect the overall inactive formation of high-mass stars, however, when the maser is excited, it exhibits comparable luminosity with regard to the other three arms.

\subsubsection{Luminosity difference of the 6.7 GHz methanol masers between on- and inter-arms} \label{subsubsec:4.2.5}
\begin{figure}[htbp!]
    \centering
    \includegraphics[width=0.5\linewidth]{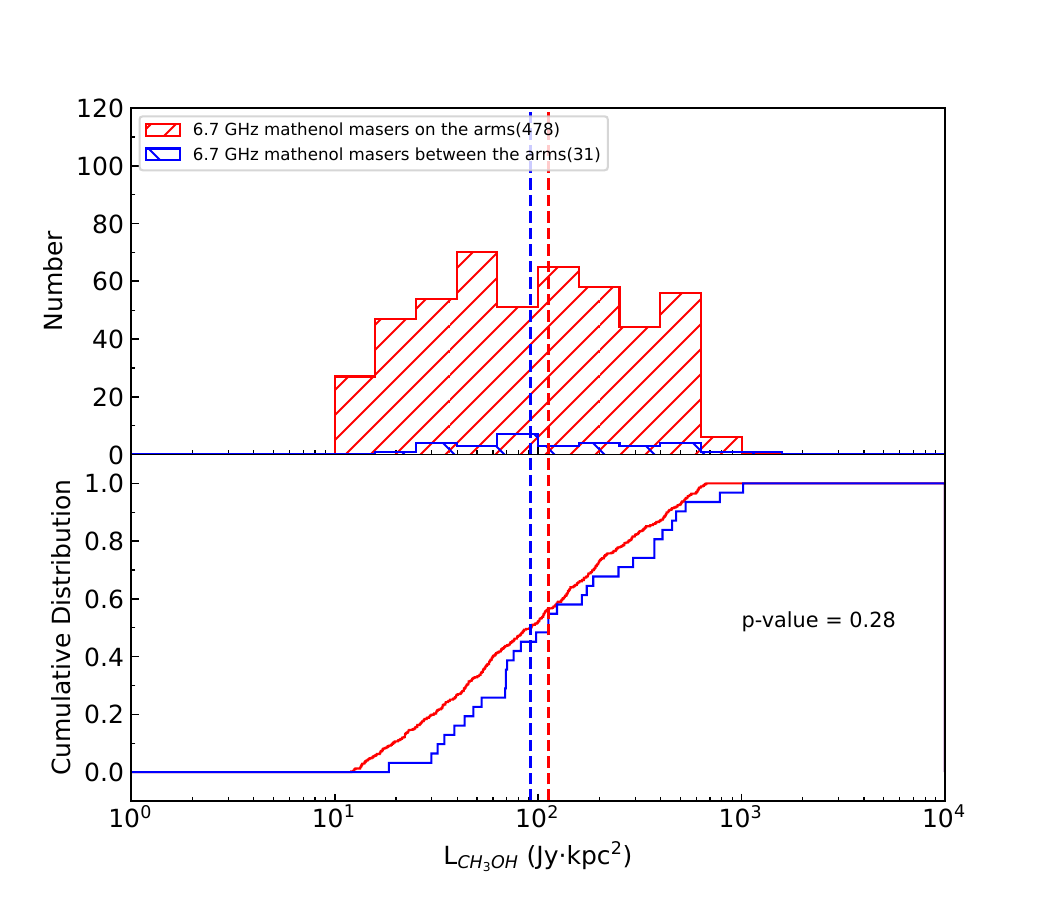}
    \caption{Comparison of the luminosity of the 6.7 GHz methanol masers from the on- and inter-arms. The legend in the upper left corner of the figure above describes the names of each arm and the number of sources located in each arm. The vertical dashed lines represent the median luminosity from the on- and inter-arms. The p-value from the K-S tests is shown in the plot.}
    \label{fig L on and between the arms_del_revised}
\end{figure}

Figure \ref{fig Overlaying} shows that the majority of the 6.7 GHz masers are located in the spiral arm regions, with some appearing in the regions between the spiral arms (i.e., inter-arms). The HMSFRs locating in the inter-arms has typically simple and less-clustered environments, making them easier to be observed and explored, thus providing valuable insights into the high-mass star formation \citep{2019ApJ...871..198C}. During comparing the differences in luminosity of the 6.7 GHz methanol masers between on- and inter-arms, as shown in Figure \ref{fig L on and between the arms_del_revised}, we only considered the sources after correcting the observing effect (see Section \ref{subsubsec:4.1.1}). Moreover, we excluded sources with excessively high or low luminosity and retained only those within the range of $\mu$ $\pm$ $\sigma$. Notably, the spiral arm sources are deemed from the only four main spiral arms without taking into account the arms of smaller segments, such as the Local arm, and the ``Unknown'' sources are deemed to be situated in the inter-arm regions (see Section \ref{subsec:3.1}). Our results show no obvious difference detected in the 6.7 GHz methanol maser luminosity between on- and inter-arms, suggesting that the physical conditions responsible for their excitation are comparable in these regions. It further indicates that the physical environments of HMSFRs which harbor the 6.7 GHz methanol masers are similar between on-arm and inter-arm regions.

\subsection{Molecular cloud environment of 6.7 GHz methanol masers traced by the ammonia gas} \label{subsec:4.3}
\subsubsection{Variation as the galactocentric distance} \label{subsubsec:4.3.1}
\begin{figure}[htbp!]
\centering
\begin{tabular}{@{}cc@{}}
\includegraphics[width=0.4\linewidth]{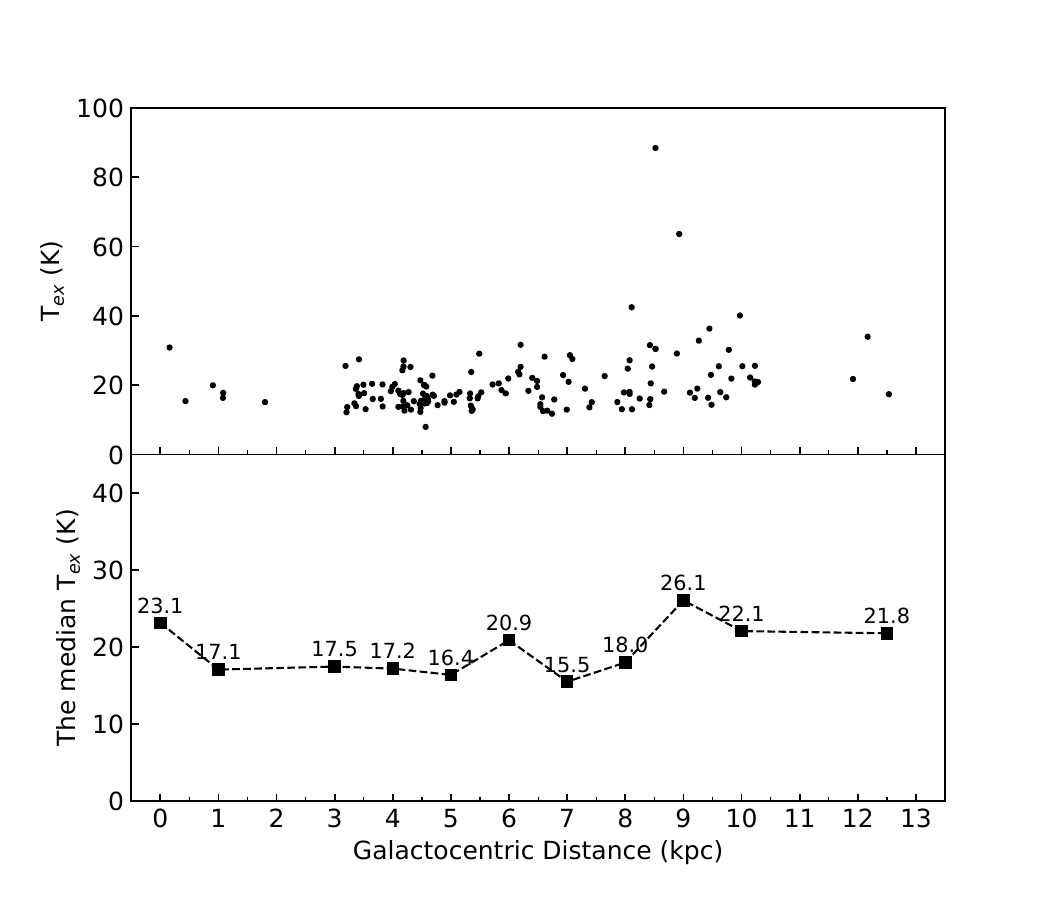}
\includegraphics[width=0.4\linewidth]{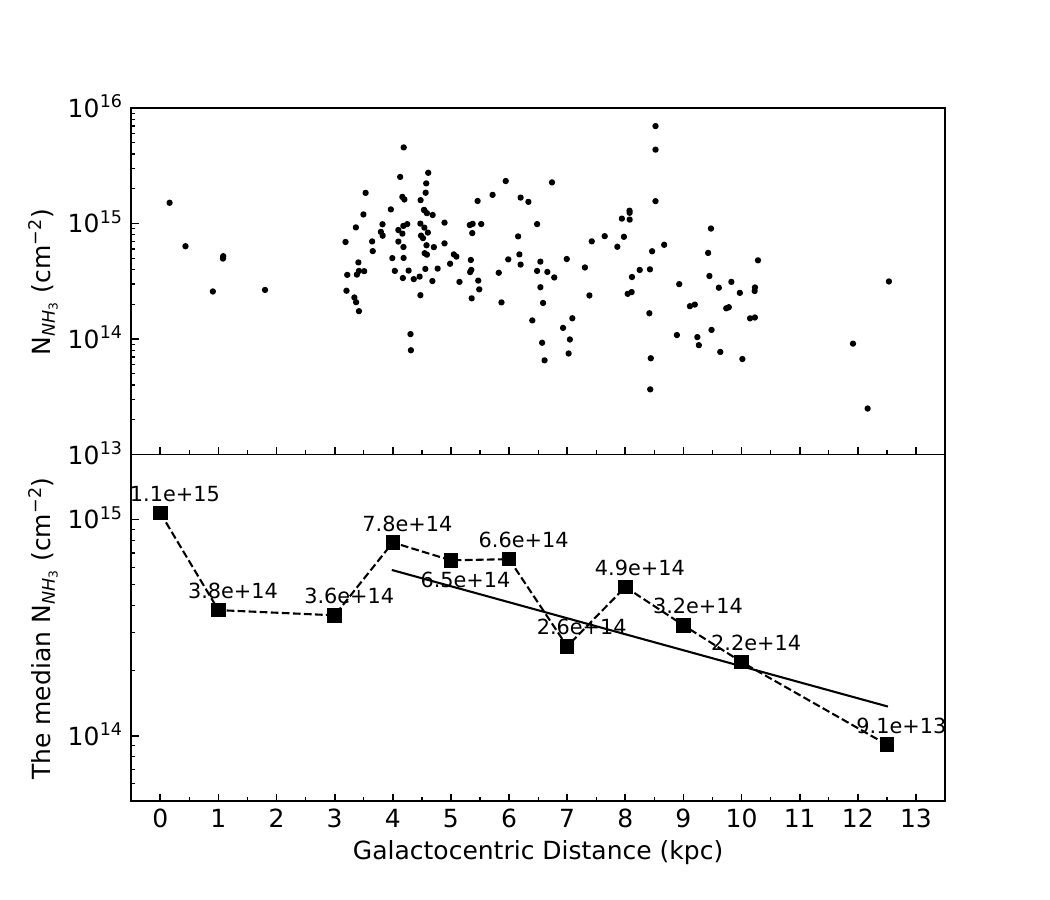}
\end{tabular}
\caption{Left: ammonia gas temperature as a function of galactocentric distance. Right: ammonia gas column density as a function of galactocentric distance.}
\label{fig Tex N and D_GC}
\end{figure}

Obtaining variations in temperature and density of molecular clouds across the galaxy scale is crucial for comprehending the structure and evolution of the Milky Way. In Figure \ref{fig Tex N and D_GC}, we present the variations of the excitation temperature and column density derived from ammonia lines observed with TMRT (see Section \ref{subsec:2.2}) as galactocentric distances. The distributions in the left panel of Figure \ref{fig Tex N and D_GC} show that the ammonia excitation temperatures are relatively stable at $\sim20$ K, without a significant variation trend as galactocentric distances, which is consistent with the ammonia observations made by \citet{2018A&A...609A.125W}. This reveals that there is no substantial difference of the gas temperature between star forming regions inside and outside of the Milky Way. It is mainly because that the gas temperature is more likely linked to the local environment of star forming regions, rather than the overall structure of the Galaxy. 

Regarding the column density, the scatter plot in the right-upper panel of Figure \ref{fig Tex N and D_GC} indicates that the ammonia sources are most dense at 4-5 kpc from the Galactic centre. This is consistent with the highest amounts of ammonia sourcs detected at 4.2 kpc and 5.7 kpc from the Galactocentre found by \citet{2012MNRAS.426.1972P}, and the presence of two peaks in the distribution of ATLASGAL sources at galactocentric distances of 4.5 kpc and 6 kpc, respectively, found by \citet{2012A&A...544A.146W}.
There is a downward trend in column density as galactocentric distances, which becomes more apparent when considering the median of these data points shown in the right-bottom panel. On the Galactic scale, the column density is generally high in the central region of the Milky Way, and gradually declines as the galactocentric distances. Additionally, we observed that the column density in the heads of the spiral arms is greater than that in the tails of the spiral arms when examining only the spiral arm regions between 3-10 kpc (i.e., excluding the central regions within 3 kpc or outer regions of $>$ 10 kpc due to the relatively small number of sources in these regions). Combining this result with the remarkable decrease of 6.7 GHz methanol maser luminosity as the galactocentric distances (see Section \ref{subsubsec:4.2.3}), it can be concluded that the heads of the spiral arms generally exhibit intense high-mass star formation activity compared to the outer region of the Milky Way. \citet{2011ApJ...741..110D} reported a decrease in the column density and abundance of ammonia with increasing galactocentric distance, and the inner Galactic sources are about seven times larger than the outer ones. This result is consistent with our findings.

\subsubsection{Differences in four main spiral arms} \label{subsubsec:4.3.2}
\begin{figure}[htbp!]
    \centering
    \includegraphics[width=0.40\textwidth]{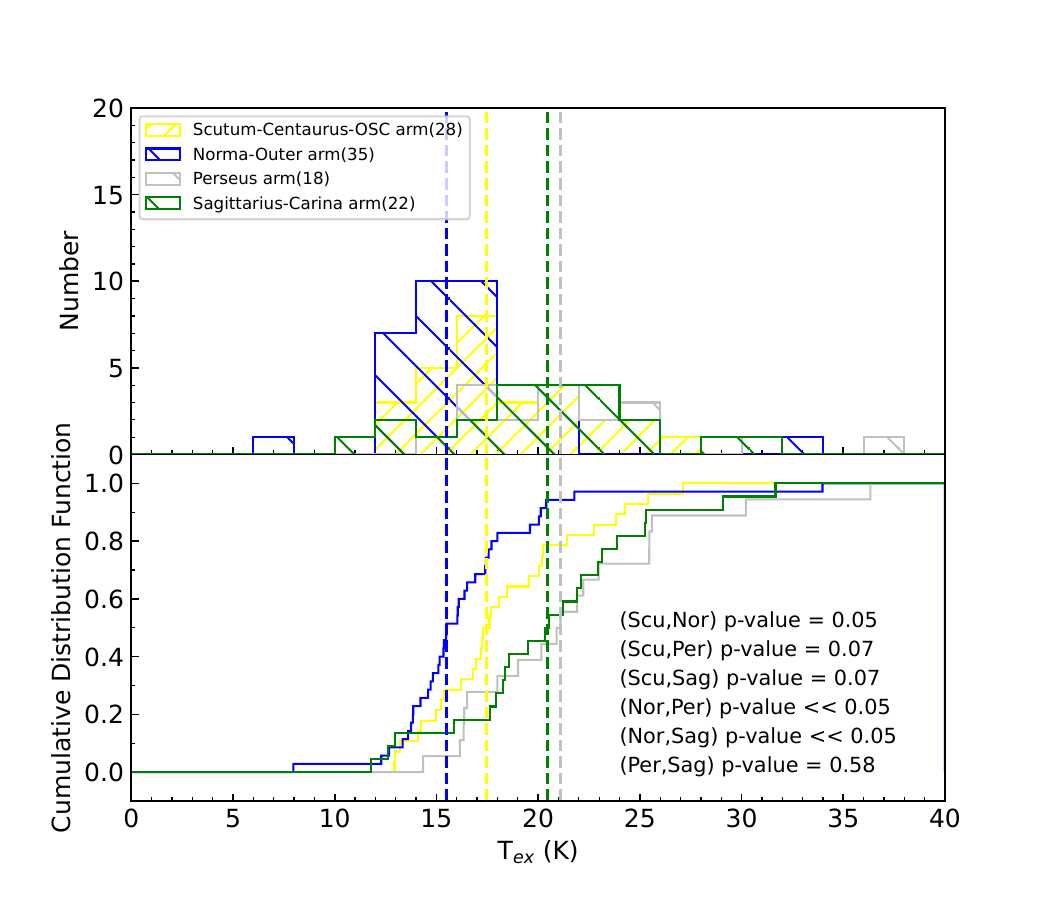}
    \includegraphics[width=0.40\textwidth]{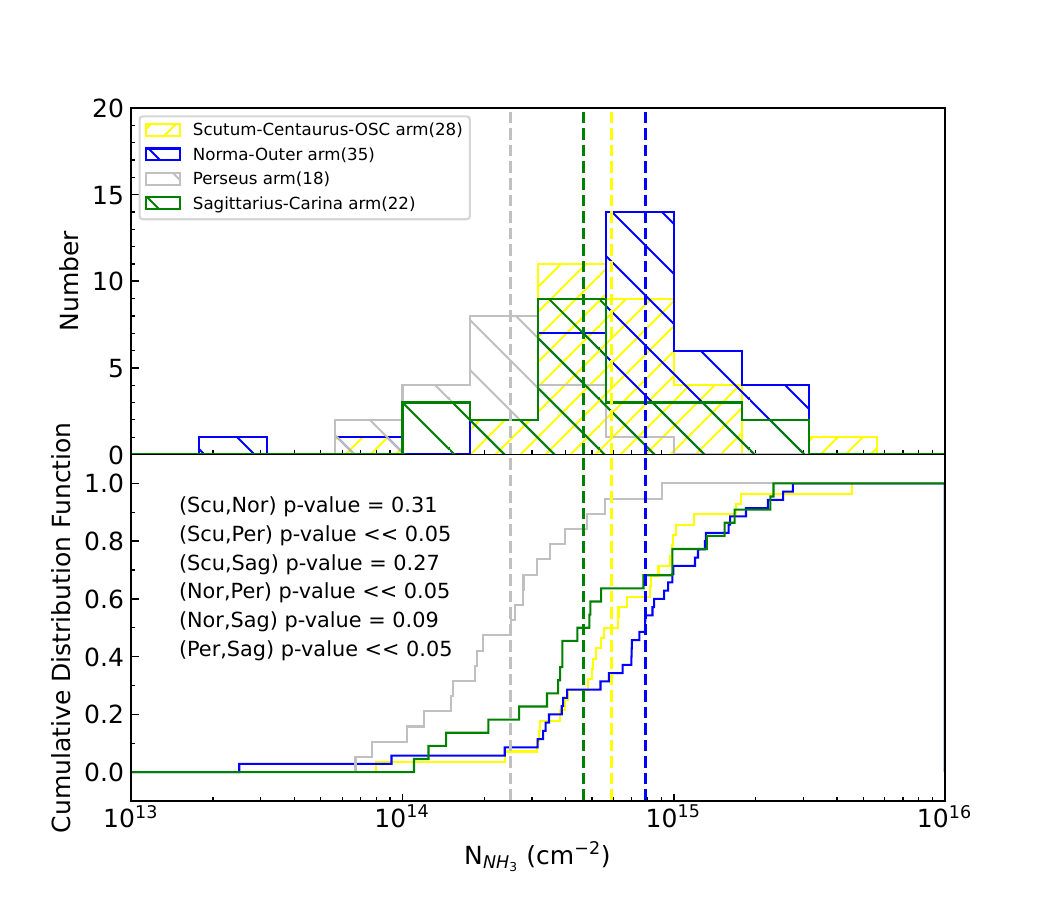}
    \caption{Left: ammonia gas temperature distribution at four spiral arms. Right: ammonia column density distribution at four spiral arms. The dashed lines of various colors represent the median values of the corresponding distributions in different arms. The p-values from the K-S tests are presented in each panel.}
    \label{fig Tex N-four arms}
\end{figure}

In Section \ref{subsubsec:4.2.4}, we investigate the variation in luminosity of the 6.7 GHz methanol masers in the four main spiral arms of the Milky Way and discover that the median maser luminosity is the bright in the Perseus arm, although the number of existed methanol masers is lowest therein. To further explore the differences among the four main spiral arms, we investigated the variations in ammonia excitation temperature and column density in these arms. Figure \ref{fig Tex N-four arms} displays the statistical histograms of temperature and column density of ammonia gas across the four spiral arms. The left panel of Figure \ref{fig Tex N-four arms} indicates that there are obvious differences in the gas temperatures among the four spiral arms, with the Perseus arm being the highest. While the right panel of Figure \ref{fig Tex N-four arms} reveals that the gas column density of the Perseus arm is significantly lower than those of the other three arms. This implies that the gas in the Perseus arm is less abundant than in the other three spiral arms. Therefore, the overall high-mass star formation should be comparatively less active in the Perseus arm compared to the other arms, which is consistent with the statistical results from the maser luminosity described in Section \ref{subsubsec:4.2.4}. \citet{2013ApJ...775...79Z} discovered a lack of the high-mass star formation activity in the range of 50$^{\circ}$ $\leq$ $l$ $\leq$ 80$^{\circ}$ in the Perseus arm due to absence of H$_2$O maser sources in those regions. Furthermore, \citet{2019ApJ...885..131R} also suggested that the Perseus arm may not be dominated by high-mass star formation activity from the statistical analysis for the HMSFRs with maser trigonometric parallax measurements. Both studies are consistent with our statistical results. It can be suggested that the low gas column density of the Perseus arm leads to a small number of masers, whereas the high gas temperatures of the HMSFRs in Perseus arm derived from ammonia observations (see above) possibly make the masers have high luminosity.

\subsubsection{Differences between on- and inter-arms} \label{subsubsec:4.3.3}
\begin{figure}[htbp!]
    \centering
    \includegraphics[width=0.40\textwidth]{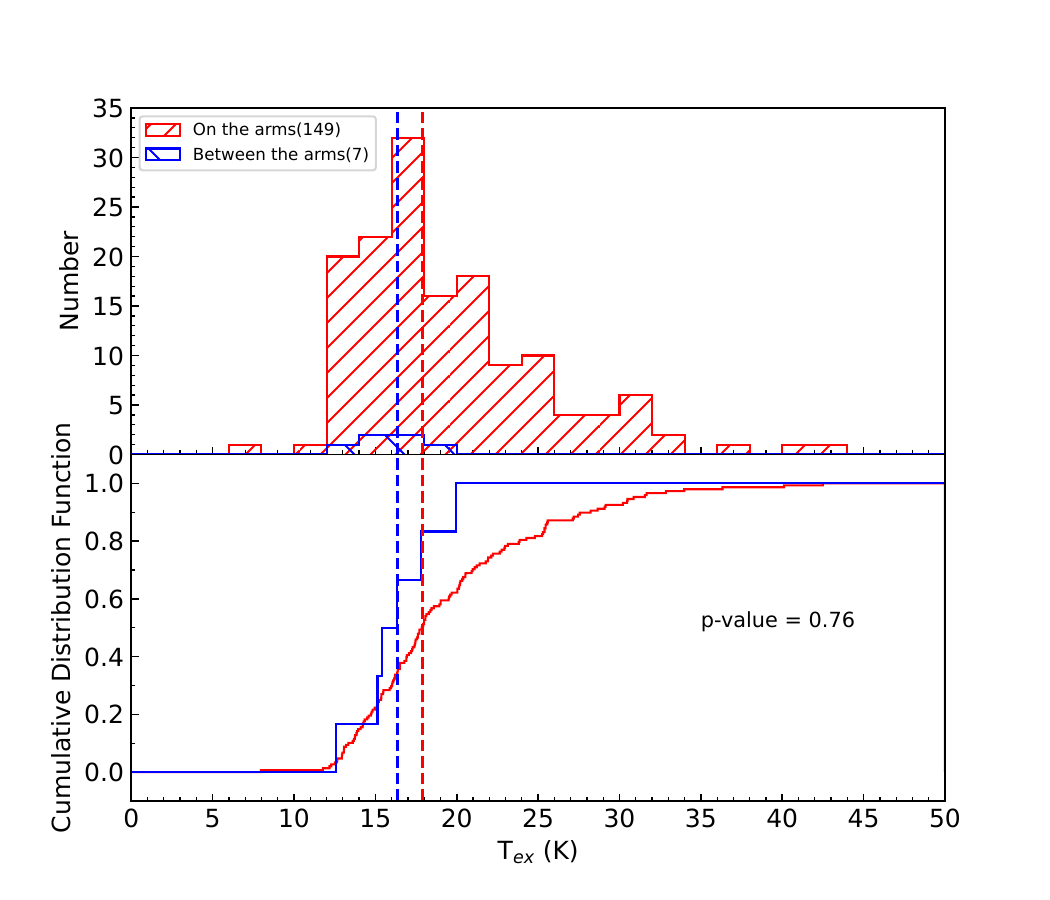}
    \includegraphics[width=0.40\textwidth]{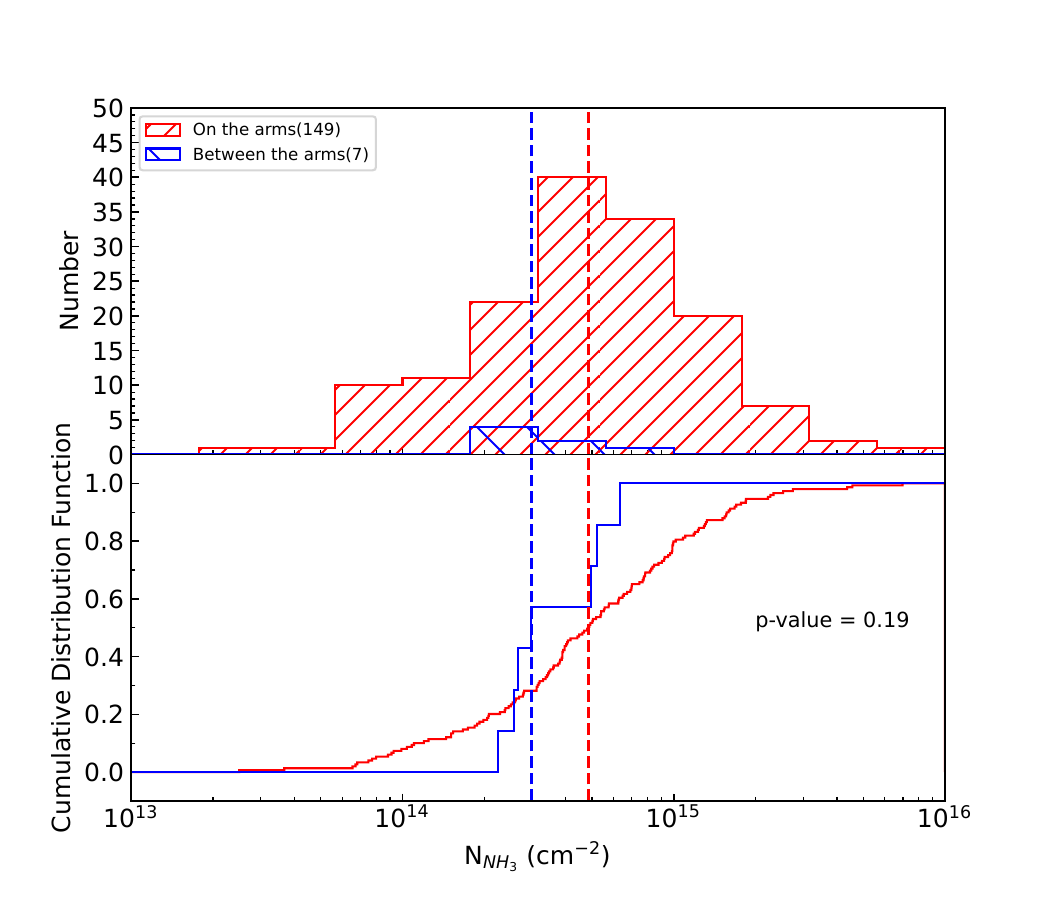}\\
    \includegraphics[width=0.40\textwidth]{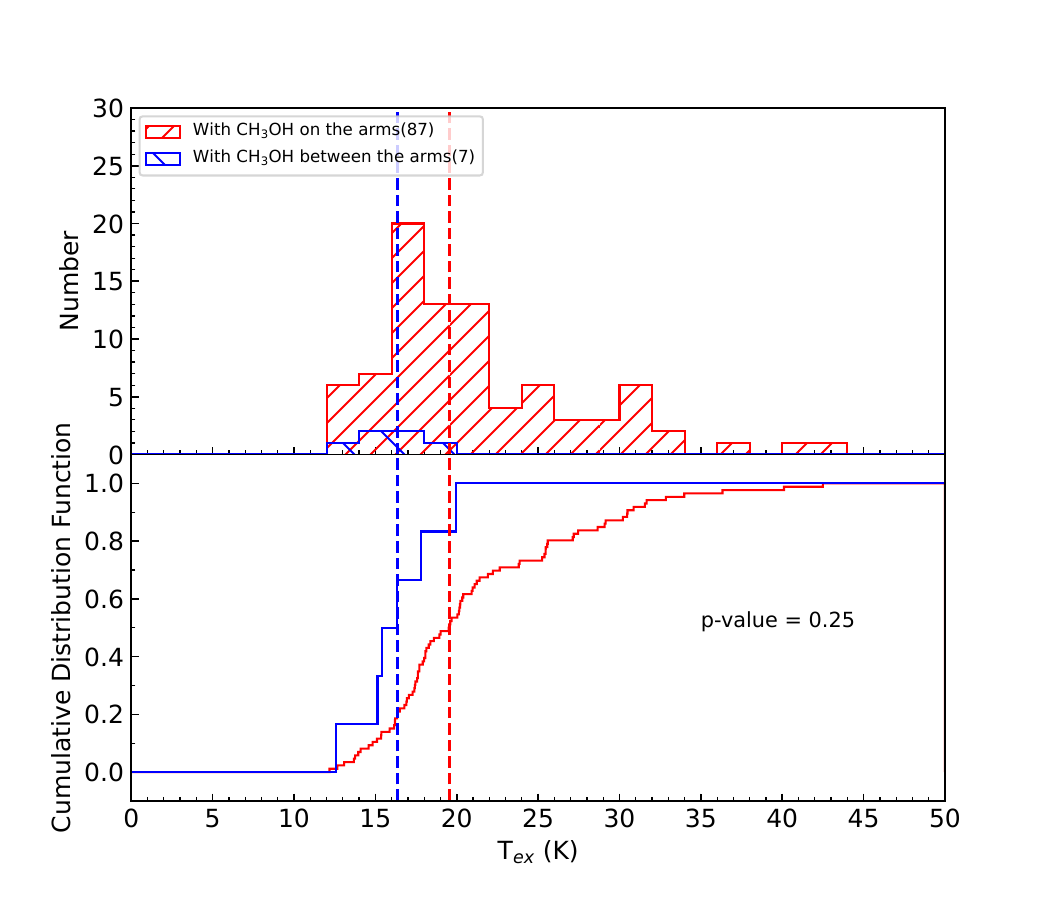}
    \includegraphics[width=0.40\textwidth]{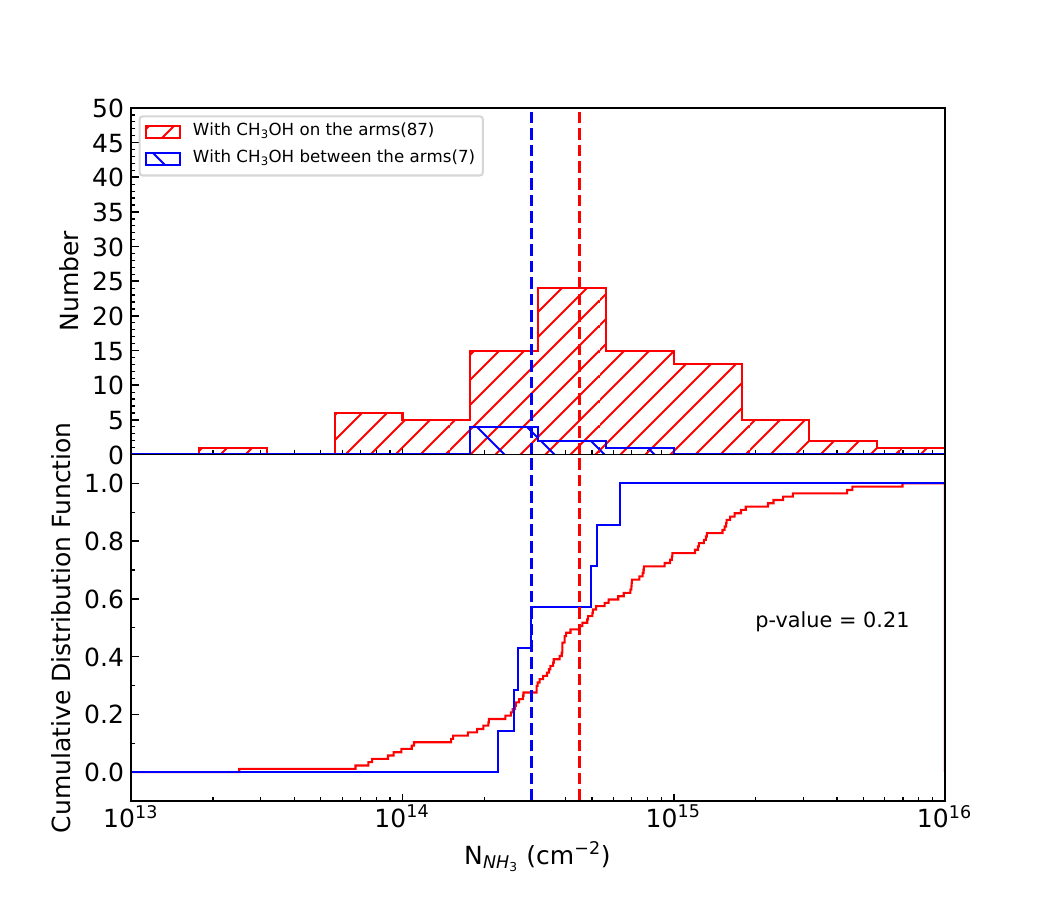}
    \caption{Upper two panels: distributions of the ammonia gas temperatures (left) and ammonia gas column density (right) in the on- and inter-arm regions for all the detected ammonia sources. Bottom two panels: same as upper two panels, but only for the ammonia sources with 6.7 GHz methanol masers.  The dashed red and blue lines represent the corresponding median of their distributions. The p-values from the K-S tests are presented in each panel.}
    \label{fig Tex N on and between the arms}
\end{figure}

We have derived that the luminosities of 6.7 GHz methanol masers are almost identical in the on- and inter-arm regions (see Section \ref{subsubsec:4.2.5}), suggesting that the physical conditions in the HMSFRs associated with methanol masers are comparable in these regions. To investigate this further, we made the statistical histograms of the temperature and column density of ammonia gas toward on- and inter-arms, as shown in Figure \ref{fig Tex N on and between the arms}. The left-upper panel of Figure \ref{fig Tex N on and between the arms} illustrates that the gas temperatures of star forming regions in the on-arm are slightly higher than that in the inter-arm. In addition, the right-upper panel of Figure \ref{fig Tex N on and between the arms} demonstrates that the gas column density in the on-arm regions is also higher than that in the inter-arm regions. Further, in order to more accurately compare the differences in the physical environment of the HMSFRs between on- and inter-arms, we made the same analysis only for the ammonia sources associated with the 6.7 GHz methanol masers, as shown in the two bottom panels of Figure \ref{fig Tex N on and between the arms}. They show a similar statistical results as that for all ammonia sources shown in the two upper panels of this figure.  
This suggests that the material of HMSFRs in the inter-arm regions should be less dense than that on the arms, making it less conducive to achieve high-mass star formation activity.

\subsection{Relationship between ammonia and 6.7 GHz methanol masers} \label{subsec:4.4}
An important approach to understanding the formation of high-mass stars is to investigate the physical condition differences between HMSRFs associated with different tracers. In this section, we focus on the relationship between the tracers of ammonia and 6.7 GHz methanol maser in HMSFRs.

\subsubsection{Difference of ammonia with or without 6.7 GHz methanol maser} \label{subsubsec:4.4.1}
\begin{figure}[htbp!]
    \centering
    \includegraphics[width=0.40\textwidth]{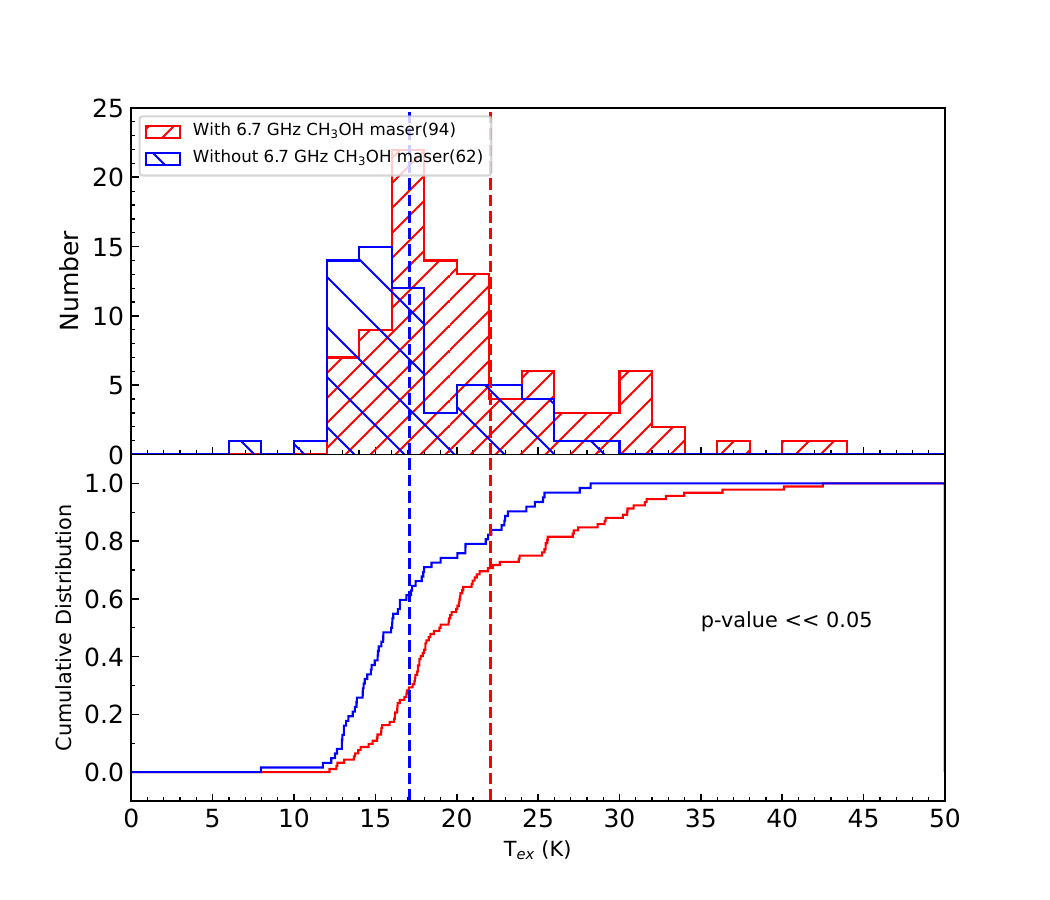}
    \includegraphics[width=0.40\textwidth]{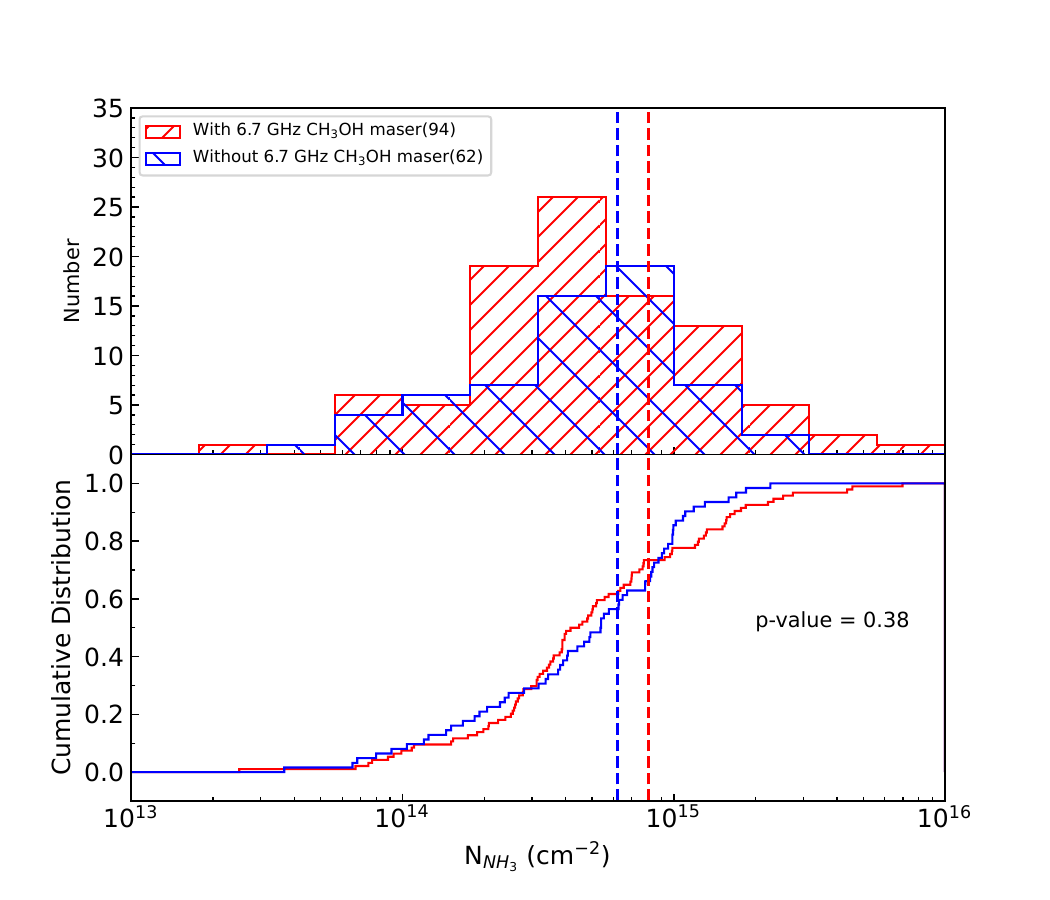}
    \caption{Left: distribution of the ammonia gas  temperatures in sources with and without 6.7 GHz methanol masers. Right: distribution of the ammonia gas column densities in sources with and without 6.7 GHz methanol masers. The dashed red and blue lines represent the means of the corresponding distributions. The p-values from the K-S tests are presented in each panel.}
    \label{fig Tex N wiht and without CH3OH}
\end{figure}

The left panel of Figure \ref{fig Tex N wiht and without CH3OH} illustrates the difference of ammonia excitation temperatures between sources with and without the 6.7 GHz methanol masers. It can be clearly seen that the sources associated with methanol masers generally have high gas temperatures than those without masers. This indicates that the 6.7 GHz methanol maser is likely produced in warmer regions. 
\citet{2020MNRAS.493.2015J} obtained the temperatures of the clumps from the SED fitting and discovered that protostellar clumps hosting class II methanol masers are hotter than the protostellar clumps without methanol masers, which is consistent with our findings. Similar result was found by \citet{2017MNRAS.470.1462L}, with HOPs ammonia sources associated with methanol masers and water masers are warmer and have larger line widths compared to those without maser emission. We believe these results are reasonable because as a molecular cloud evolves, its ambient temperature becomes warmer. Consequently, the higher excitation temperature of molecular clouds hosting methanol masers suggests that they are linked to a later evolutionary phase of massive star formation compared to those without methanol masers.
The right panel compares the ammonia column densities in regions with and without the methanol masers. It shows that there is no obvious differences of ammonia column density between sources with and without the methanol masers, although it seems that the masers can be detected toward the regions with higher gas densities. 

\subsubsection{Difference of 6.7 GHz methanol maser with or without ammonia} \label{subsubsec:4.4.2}
\begin{figure}[htbp!]
    \centering
    \includegraphics[width=0.40\textwidth]{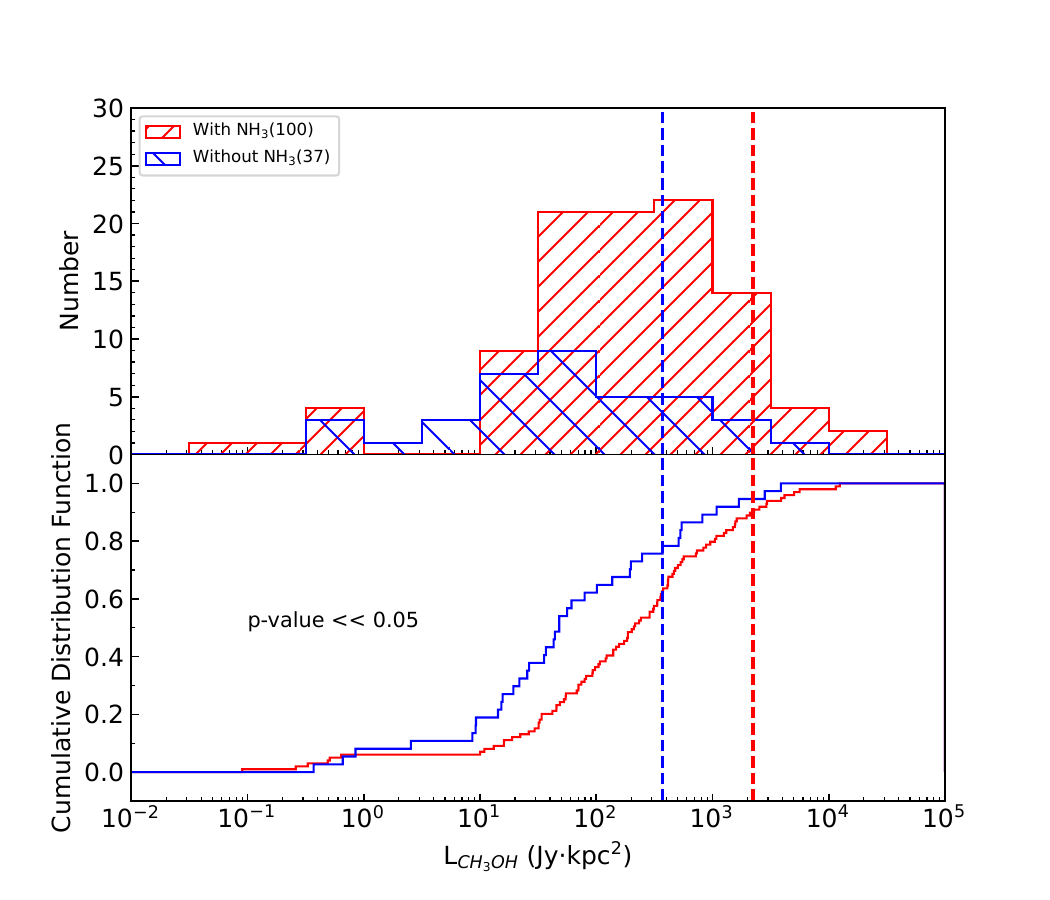}
    \caption{Luminosity distribution of 6.7 GHz methanol maser in sources with or without ammonia emission. The dashed red and blue lines represent the means of the corresponding distributions. The p-value from the K-S tests is shown in the plot.}
    \label{fig L with and without NH3}
\end{figure}

We examined the difference in 6.7 GHz methanol maser luminosity between sources with and without ammonia emission, as shown in Figure \ref{fig L with and without NH3}. The statistical result shows that the luminosity of the 6.7 GHz methanol maser is generally greater in sources with ammonia emission than those without. This suggests that the 6.7 GHz methanol maser source with ammonia emission is closer to the later evolutionary stage of massive star formation than those without ammonia emission. Additionally, the methanol maser luminosity in regions with ammonia detection is usually greater than 10 Jy kpc$^2$. Therefore, the detection of ammonia is more inclined toward regions with active methanol maser emission.

\subsubsection{Relationship between the luminosity and flux of methanol and ammonia}\label{subsubsec:4.4.3}
\begin{figure}[htbp!]
    \centering
    \includegraphics[width=0.40\textwidth]{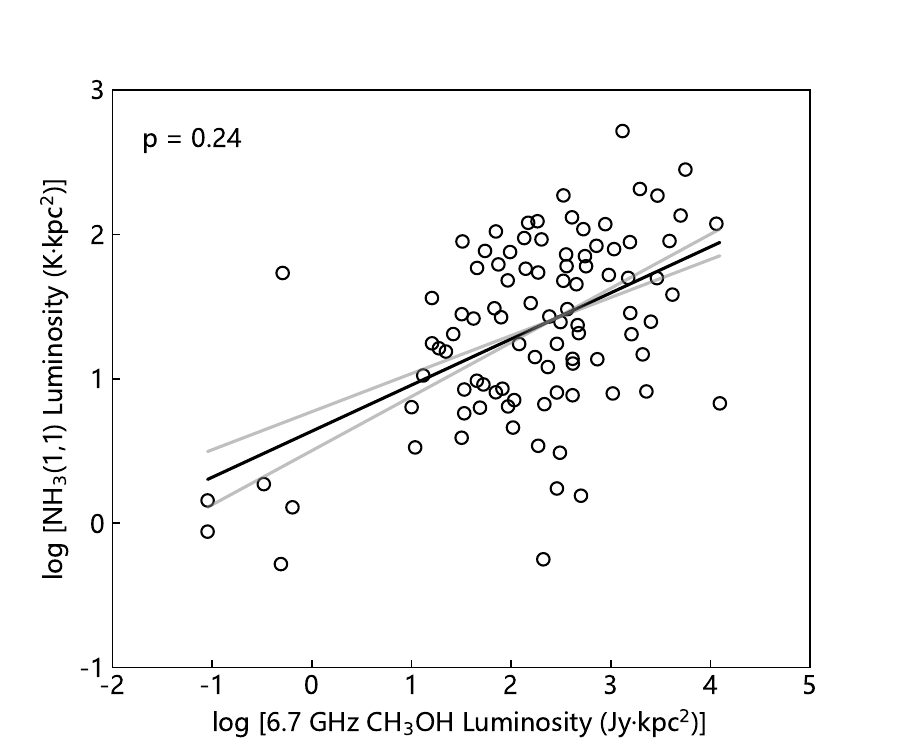}
    \includegraphics[width=0.40\textwidth]{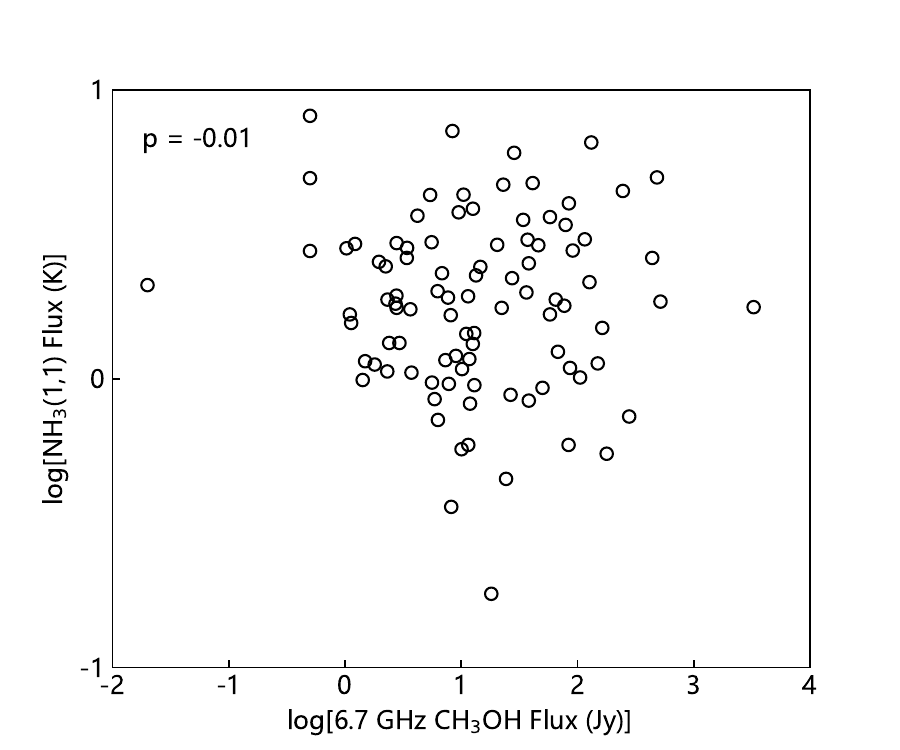}
    \caption{Correlation analysis of luminosity and flux for the 6.7 GHz methanol maser and NH$_3$ (1, 1) emission. The black line is the linear fit to log-log correlation of their luminosities, and the two gray lines are the error boundary of that fit line. The p-value in the left-upper corner is the Pearson correlation coefficient.}
    \label{fig L(CH3OH NH3)_new}
\end{figure}

Figure \ref{fig L(CH3OH NH3)_new} shows the correlation analysis of luminosities between the 6.7 GHz methanol maser and NH$_3$ (1, 1) emission. The fitting equation for the left panel of Figure \ref{fig L(CH3OH NH3)_new} obtained by the least square method can be expressed as
\begin{equation}
    log(L_{NH_3(1, 1)}) = 0.32[0.06]log(L_{CH_3OH}) + 0.64[0.13],
\end{equation}
where $L_{NH_3(1, 1)}$ is the luminosity of the NH$_3$ (1, 1) (K kpc$^2$) and $L_{CH_3OH}$ is the luminosity of the 6.7 GHz methanol maser (Jy kpc$^2$). The number in the brackets represent the standard errors. It seems that there is a weak positive luminosity correlation between the two tracers, with a low Pearson correlation coefficient (p=0.24). When the distance of the sources is not taken into account and just their flux densities are compared, as shown in the right panel of Figure \ref{fig L(CH3OH NH3)_new}, there is no correlation between them.

\section{Summary} \label{sec:5}
We compiled an updated catalog of 6.7 GHz methanol masers consisting of 1092 sources throughout the Milky Way and conducted a statistical analysis for their luminosity and spatial distributions. Additionally, we made ammonia observation toward 214 star forming regions using TMRT. We derived the excitation temperature and column density toward regions with ammonia detection, and made statistical analysis to the physical environments traced by the ammonia gas, as well as linked them with the statistical findings of the 6.7 GHz methanol masers. The main findings of our study are as follows:

\begin{enumerate}
\item The 6.7 GHz methanol masers cover the entire Galactic plane. The most dense concentration of 6.7 GHz methanol maser sources is located at distances between 4--7 kpc from the Galactic center, indicating that high-mass star formation is most active in these regions.
\item It is observed that the luminosity of 6.7 GHz methanol maser and the column density of ammonia gas decreases with galactocentric distances. This suggests that high-mass star formation activity is more intense and abundant in the inner Milky Way ($<$ 8.15 kpc) compared to the outer Milky Way ($>$ 8.15 kpc). The excitation temperature of ammonia gas has no obvious trend with the increase of galacticocentric distance. This reveals that the gas temperature is more likely to be related to the local environment of the star-forming region than to the large-scale structure of the Milky Way.
\item Compared to the other three arms, the gas column density of the Perseus arm is lower, indicating that it may not be dominated by HMSFRs, which may be the reason for the scarcity of 6.7 GHz methanol masers detected in the Perseus arm. However, the high gas excitation temperature of the Perseus arm makes makes the maser luminosity large in this arm.
\item No obvious difference detected in the 6.7 GHz methanol maser luminosity between on- and inter-arms, suggesting that the physical environments of HMSFRs which harbor the 6.7 GHz methanol masers are similar between on-arm and inter-arm regions. The gas column density in the inter-arm regions is smaller than that on arm regions, indicating that the inter-arm material is less dense and therefore less conducive to high-mass star formation.
\item In regions where 6.7 GHz methanol maser is present, the excitation temperature and column density of ammonia gas are higher compared to regions without it. Conversely, the luminosity of the 6.7 GHz methanol maser is higher in regions with ammonia emission, indicating a correlation between ammonia and the 6.7 GHz methanol maser. 
\end{enumerate}

\acknowledgments
This work is supported by the National Key R\&D program of China (2022YFA1603102, 2022YFA1603100) and the National Natural Science Foundation of China (11873002, 12011530065, 11590781). We are grateful to the TMRT for providing the observational data for this paper, as well as to the operators for their support and assistance during the observations. X.C. thanks the Guangdong Province Universities and Colleges Pearl River Scholar Funded Scheme (2019).

\bibliography{6.7GHz_methanol}{}
\bibliographystyle{aasjournal}

\appendix
\section{Updated catalog of the 6.7 GHz methanol masers\label{appendix/methanol}}

\setcounter{table}{0}  
\renewcommand{\thetable}{A\arabic{table}}

\begin{longrotatetable}
\begin{deluxetable*}{ccrrrrrccrrrcrrrl}
\tablecaption{Updated catalog of 6.7 GHz methanol masers\label{table/mathanol}}
\tabletypesize{\scriptsize}
\setlength{\tabcolsep}{0.03in}
\tablehead{
\colhead{Name} & \colhead{R.A.(J2000)} & \colhead{Dec.(J2000)} & \colhead{V$_l$} & \colhead{V$_u$} & \colhead{V$_p$} & \colhead{S$_p$} & \colhead{Para} & \colhead{+/-} & \colhead{Dist} & \colhead{+/-} & \colhead{Prob} & \colhead{Arm} & \colhead{D$_{GC}$} & \colhead{SA} & \colhead{L} & \colhead{Ref}\\ 
\colhead{($^{\circ}$ $^{\circ}$)} & \colhead{(h m s)} & \colhead{($^{\circ}$ $^{\prime}$ $^{\prime\prime}$)} & \colhead{(km s$^{-1}$)} & \colhead{(km s$^{-1}$)} & \colhead{(km s$^{-1}$)} & \colhead{(Jy)} & \colhead{(mas)} & \colhead{(mas)} & \colhead{(kpc)} & \colhead{(kpc)} & \colhead{} & \colhead{} & \colhead{(kpc)} & \colhead{($^{\circ}$)} & \colhead{(Jy kpc$^2$)} & \colhead{}
}
\decimalcolnumbers
\startdata
G000.092$-$00.663       & 17 48 25.90 & $-$29 12 05.9 & 10.0   & 25.0   & 23.5   & 24.80   &       &       & 2.82  & 0.22  & 0.96 & ScN & 5.33  & 179.95 & 197.22    & MMB     \\
G000.167$-$00.446       & 17 47 45.46 & $-$29 01 29.3 & 9.5    & 17.0   & 13.8   & 4.44    &       &       & 2.81  & 0.23  & 0.95 & ScN & 5.34  & 179.91 & 35.06     & MMB     \\
G000.212$-$00.001       & 17 46 07.63 & $-$28 45 20.9 & 41.0   & 50.5   & 49.3   & 3.47    &       &       & 11.10 & 0.24  & 0.85 & 3kF & 2.95  & 0.80   & 427.54    & MMB     \\
G000.315$-$00.201       & 17 47 09.13 & $-$28 46 15.7 & 14.0   & 27.0   & 19.4   & 72.16   & 0.342 & 0.042 & 2.92  & 0.36  & 1.00 & ScN & 5.23  & 179.82 & 615.27    & MMB;R19 \\
G000.316$-$00.201       & 17 47 09.33 & $-$28 46 16.0 & 20.0   & 22.0   & 21.0   & 0.60    &       &       & 2.84  & 0.23  & 0.66 & ScN & 5.31  & 179.83 & 4.84      & MMB     \\
G000.376$+$00.040       & 17 46 21.41 & $-$28 35 40.0 & 35.0   & 40.0   & 37.0   & 2.32    & 0.125 & 0.047 & 8.00  & 3.01  & 1.00 & 3kF & 0.16  & 160.73 & 148.48    & MMB;R19 \\
G000.390$-$00.030       & 17 46 41.12 & $-$28 37 05.5 & 22.0   & 31.0   & 28.7   & 5.80    &       &       & 12.48 & 0.25  & 0.34 & SgF & 4.33  & 1.12   & 903.35    & P05     \\
G000.409$-$00.504       & 17 48 33.48 & $-$28 50 52.5 & 24.5   & 27.0   & 25.3   & 2.77    &       &       & 2.83  & 0.23  & 1.00 & ScN & 5.32  & 179.78 & 22.18     & MMB     \\
... & ... & ... & ... & ... & ... & ... & ... & ... & ... & ... & ... & ... & ... & ... & ... & ... \\    
\enddata
\tablecomments{Columns (1)--(3): source name, and source positions in equatorial coordinates. Columns (4)--(6): the lower and upper velocities and the peak velocity of the 6.7 GHz methanol maser spectra. Column (7): the maser peak flux density. Columns (8) and (9): parallax and its error. Columns (10) and (11): distance and its error. Column (12): the probability of being in the spiral arm. Column (13): the spiral arm of the source -- ScN (Scutum near), ScF (Scutum far), CtN (Centaurus near), CtF (Centaurus far), Nor (Norma arm), N1N (Norma arm Q1 near), N1F (Norma arm Q1 far), N4N (Norma arm Q4 near), N4F (Norma arm Q4 far), Out (Outer arm), Per (Perseus arm), SgN (Sagittarius near), SgF (Sagittarius far), CrN (Carina near side of tangent point), CrF (Carina far side of tangent point), 3kN (near 3-kpc arm), 3kF (far 3-kpc arm), Loc (local arm), Los (local spur), AqS (Aquila spur), Con (Connecting arm), GC (Galactic center region), ... (Unknown). Column (14): Galactocentric distance. Column(15): spiral angle. Column(16): maser luminosity. Column (17): References -- MMB (the parkes MMB survey), P05 (Pestalozzi et al. 2005), P07 (Pandian et al. 2007), X08 (Xu et al. 2008), S12 (Szymczak et al. 2012), O14 (Olmi et al. 2014), Y17 (Yang et al. 2017), Y19 (Yang et al. 2019), O19 (Ouyang et al. 2019), R19 (Reid et al. 2019).\\
(This table is available in its entirety in machine-readable form.)}
\end{deluxetable*}
\end{longrotatetable}

\end{document}